\documentclass[fleqn,usenatbib,useAMS]{mnras}

\usepackage{natbib}
\usepackage{cleveref}
\creflabelformat{equation}{#2#1#3}
\usepackage{ulem}
\usepackage{ragged2e}
\usepackage{graphicx}
\usepackage{color}
\usepackage{txfonts}

\usepackage{wrapfig}
\usepackage{subcaption}
\newcommand{\msun}{\ensuremath{\mathrm{M}_{\odot}}}   
\newcommand{\mini}{\ensuremath{M_{\rm ini}}}                         

\newcommand{\angstrom}{\mbox{\normalfont\AA}}

\newcommand{\vcrit}{\ensuremath{\upsilon_{\rm crit}}}                         
\newcommand{\vini}{\ensuremath{\upsilon_{\rm ini}}}                         

\newcommand{\teff}{\ensuremath{\mathit{T}_{\rm eff}}}                

\newcommand{\mmax}{\ensuremath{M_{\mathrm{max}}}}  
\newcommand{\mmin}{\ensuremath{M_{\mathrm{min}}}}  

\newcommand{\fesc}{\ensuremath{f_{\mathrm{esc}}}} 

\newcommand{\Qi}{\ensuremath{Q_{\mathrm{i}}}}   
\newcommand{\QH}{\ensuremath{Q_{\mathrm{H}}}}   
\newcommand{\QHeI}{\ensuremath{Q_{\mathrm{HeI}}}}   
\newcommand{\QHeII}{\ensuremath{Q_{\mathrm{HeII}}}}   
\newcommand{\Qrot}{\ensuremath{Q_{\mathrm{rot}}}}   
\newcommand{\Qnorot}{\ensuremath{Q_{\mathrm{no rot}}}}   
\newcommand{\Ni}{\ensuremath{N_{\mathrm{i}}}}   
\newcommand{\NH}{\ensuremath{N_{\mathrm{H}}}} 
\newcommand{\NHeI}{\ensuremath{N_{\mathrm{HeI}}}} 
\newcommand{\NHeII}{\ensuremath{N_{\mathrm{HeII}}}} 
\newcommand{\Npop}{\ensuremath{N_{\mathrm{pop}}}}   

\definecolor{green}{rgb}{0.3,0.7,0.}

\definecolor{orange}{rgb}{0.5,0.2,0.0}

\begin{document} 



\title[Pop III ionizing photon production]{Ionizing photon production of Population III stars: effects of rotation, convection, and initial mass function}

\author[L.J.Murphy et al.]{Laura~J.~Murphy$^{1}$\thanks{E-mail: murphl25@tcd.ie}, 
Jose H. Groh$^{1}$\thanks{E-mail: jose.groh@tcd.ie},
Eoin Farrell$^{1}$,
Georges Meynet$^{2}$,
\newauthor
Sylvia Ekstr\"om$^{2}$,
Sophie Tsiatsiou$^{2}$, 
Alexander Hackett$^{3}$,
Sébastien Martinet$^{2}$
 \\
 $^{1}$School of Physics, Trinity College Dublin, the University of Dublin, College Green, Dublin\\
 $^{2}$Department of Astronomy, University of Geneva, Chemin des Maillettes 51, 1290 Versoix, Switzerland\\
 $^{3}$Institute of Astronomy, University of Cambridge, Madingley Road, Cambridge CB3 0HA, UK
 }

\date{Accepted XXX. Received YYY; in original form ZZZ}

\pubyear{2021}


 \label{firstpage}
 \pagerange{\pageref{firstpage}--\pageref{lastpage}}
\maketitle

\begin{abstract}
   {The first stars are thought to be one of the dominant sources of hydrogen reionization in the early Universe, with their high luminosities and surface temperatures expected to drive high ionizing photon production rates. In this work, we take our Geneva stellar evolution models of zero-metallicity stars and predict their production rates of photons capable to ionize H, He I and He II, based on a blackbody approximation. We present analytical fits in the range 1.7-500$\,\msun$. We then explore the impact of stellar initial mass, rotation, and convective overshooting for individual stars. We have found that ionizing photon production rates increase with increasing initial mass. For the rotational velocities considered we see changes of up to 25\% to ionizing photons produced. This varies with initial mass and ionizing photon species and reflects changes to surface properties due to rotation. We have also found that higher convective overshooting increases ionizing photon production by approximately 20\% for the change in overshooting considered here. For stellar populations, we explore how the production of ionizing photons varies as a function of the initial mass function (IMF) slope, and minimum and maximum initial masses. For a fixed population mass we have found changes of the order of 20-30\% through varying the nature of the IMF. This work presents ionizing photon production predictions for the most up to date Geneva stellar evolution models of Population III stars, and provides insight into how key evolutionary parameters impact the contribution of the first stars to reionization. }

\end{abstract}


\section{Introduction}\label{sec:intro}

The formation of the first stars a few hundred million years after the Big Bang \citep[e.g.,][]{Bromm2013FormationStars} fundamentally changed the history of the universe. The first stellar generation, commonly referred to as Population III (Pop~III) stars, was composed of metal-free, luminous hot stars that were the first sources of hydrogen ionizing photons, initiating the epoch of reionization.  It is generally accepted that the first stellar populations were a significant, and perhaps dominant, source for the reionization of hydrogen in the intergalactic medium (IGM; \citealt{Haehnelt2001,FaucherGiguere2008,FaucherGiguere2009,Becker2013,Wise2014}). Active galactic nuclei (AGNs) also contribute to reionization and are the drivers of the full reionization of He~I in the IGM \citep{BarkanaLoeb2001,FaucherGiguere2009,McQuinn2016,Worseck2016}.
The relative contribution of AGN and stars to cosmic reionization is a topic at the forefront of Astrophysics research, with ongoing observational and theoretical efforts from different groups. Our paper deals with one of the important aspects within this broader theme, which is the production of ionizing photons from Pop~III stars and how this is affected by fundamental, yet poorly constrained, stellar properties such as rotation and convection.

Previous studies have shown that Pop~III stars are expected to produce many ionizing photons since zero-metallicity stars are more luminous and hotter than stars at higher metallicities \citep{Marigo2001Zero-metallicityMass,Marigo2003,Ekstrom2008EffectsStars,Yoon2012EvolutionFields,Murphy2021}. The ionizing photon production rates for higher metallicity stars have been studied in \cite{Topping2015}. In that work, ionizing photon production rates were determined for a range of metallicities from stellar evolution model grids of \cite{Ekstrom2012GridsZ=0.014} and \cite{Georgy2013Grids0.002}, and also compared to work from \cite{Schaerer2003} on ionizing photon production at very low metallicities (Z\,=\,$10^{-5}\!-\!10^{-7}$). Predictions for ionizing photon production of Pop~III stars have been made in \cite{TumlinsonShull2000,Schaerer2002,HegerWoosley2010} and \cite{Yoon2012EvolutionFields}. Using the available evolutionary tracks of metal-free stars at that time \citep{1983Castellani,ElEid1983,Chieffi1984}, \cite{TumlinsonShull2000} used a fitting method to produce `static stellar models' and, combined with atmosphere models, predicted the ionizing photon production of the first stars. This work provided a first look at how much more efficient Pop~III stars are at producing ionizing photons than higher metallicity stars, but there was still much to be done in accurately modelling surface properties of the first stars. 

With new stellar evolution model grids of zero-metallicity stars, \cite{Schaerer2002} combined stellar evolution and atmosphere models to predict ionization rates. This was an important step, because for the first time it allowed the study of the evolution of ionizing photon production rates over the lifetime of Pop~III stars. Based on the Geneva stellar evolution models at that time, they found that redward evolution of Pop~III stars decreased ionizing photon production by a factor of two over the stellar lifetime. This shows the significant impact that stellar evolution can have on ionizing photon production, and why it is important to have updated predictions with the most recent model grids available. A new stellar evolution grid of Pop~III models was presented in \cite{HegerWoosley2010} for a fine grid of 120 non-rotating models in the mass range 10-100$\,\msun$, from which the ionizing photon production per solar mass was determined. This provided further insight on how a population of zero-metallicity stars may have contributed to reionization. The effects of rotation on the ionizing photon production of Pop~III stars was explored for the first time in \cite{Yoon2012EvolutionFields}. The models in that work also included internal magnetic fields and studied the impact of chemically homogeneous evolution on the first stars. This was a significant development in understanding how stellar properties impact ionizing photon production, and further demonstrates the importance of exploring how internal mixing impacts the ionizing photon contribution of the first stars. 

More recently, the contribution of stripped stars and binaries were investigated in \cite{Gotberg2020} for metallicities Z\,=\,0.014,\,0.006,\,0.002,\,and\,0.0002. They show that due to their hotter surface temperatures and higher luminosities, stripped stars are expected to significantly increase the contribution to ionization, however, this effect decreases at lower metallicities since these stars are already very hot and luminous as single stars. \cite{Berzin2021} has recently confirmed this, in showing that stellar populations at these metallicities (Z\,=$\,0.0002\!-\!0.014$) which include stripped stars produce significantly more ionizing photons than those without. They also showed that Pop~III stellar populations produce much more ionizing photons than higher metallicity populations, even those where stripped stars are included.

Given that it is expected that stellar initial mass will impact the ionizing photon production rate \citep{TumlinsonShull2000,Schaerer2002,Alvarez2006}, the initial mass function (IMF) will have an impact on the ionizing photon production of the first stellar populations. Much work has been done to understand the initial mass function of the first stars. The first hydrodynamical simulations (e.g. \citealt{Abel2002}) predicted preferential formation of very massive first stars ($\geq \! 100 \,\msun$; \citealt{Bromm2002TheCloud}). However, more recent simulations predict significant fragmentation \citep{Stacy2010,Clark2011TheProtostars}, the formation of binaries \citep{Turk2009}, and a wide initial mass distribution from tens to hundreds of solar masses \citep{Hirano2014,Hirano2015}. In fact, a redshift dependent IMF for Pop~III stars was proposed in \cite{Hirano2015}, where stars form on the order of hundreds of solar masses at redshifts z $\geq$ 20, while stars form on the order of tens of solar masses at lower redshifts due to relatively cool formation environments. Although research to constrain the exact distribution of the Pop~III IMF continues, there is general agreement on a top-heavy primordial IMF \protect\citep{Greif2011,StacyBromm2013,Susa2014,Hirano2014,Hirano2015,Stacy2016,Jerabkova2018,Wollenberg2020}. The IMF for solar metallicity stars is typically parameterized using the \cite{Salpeter1955} IMF and its recent revisions \citep{Scalo1986,Kroupa2001,Chabrier2003}, with a slope of $\alpha\!=\!2.35$ in the massive star regime. The slope here refers to the IMF relation $dN/dM\!\propto\! M^{-\alpha}$ which defines the number of stars of different initial masses ($\mini$) in a population. \cite{StacyBromm2013} performed cosmological simulations of Pop~III stellar systems in a range of minihalo environments, and in doing so were able to derive an IMF slope of $\alpha\!=\!0.17$. While much remains to be done to constrain the primordial IMF, this value is very useful for comparing to higher metallicity populations, which follow the Salpeter IMF slope. Such approaches have been used to study the enrichment of Population II stars \citep{Jaacks2018,Welsh2021}, and the statistics of Pop~III binaries \citep{Liu2021}. As observations of the early Universe increase with upcoming facilities such as the James Webb Space Telescope (JWST), the Wide Infrared Survey Telescope (WFIRST), and the Square Kilometre Array (SKA), we will soon have further constraints on the Pop~III IMF, and until then we need to consider a broad range of IMFs.  

It is an exciting time for research of the early Universe ahead of new observational facilities, and we must prepare for the groundbreaking detections that are expected in the coming decade. To understand the impact of the first stars on reionization, we need to have updated predictions for ionizing photon production from Pop~III stars, for a broad range of parameters. In \cite{Murphy2021} we presented a new grid of Pop~III stellar evolution models of initial masses $1.7\,\msun\!\leq\!\mini\!\leq\!120\,\msun$, with and without rotation. These models extend the GENEC model grid \citep{Ekstrom2012GridsZ=0.014,Georgy2013Grids0.002,Groh2019Grids0.0004} down to zero metallicity and have improved our understanding of rotational effects on the evolution of surface properties and metal enrichment in Pop~III stars. This grid provides opportunity to study the impact of rotation and initial mass on ionizing photon production. Since surface properties of these stars directly impact ionization, it is important to make updated predictions with the latest stellar evolution models, these new predictions are presented in this work. The paper is organised as follows. \Cref{sec:methods} describes the methods and the calculation of the ionizing photon production rate, the results of this work are presented in \Cref{sec:results}, we provide further discussion in \Cref{sec:discussion}, and present our conclusions in \Cref{sec:conclusions}.

\section{Methods}\label{sec:methods}

We employ our recent grid of Pop~III stellar models from \citet{Murphy2021}. This grid extends the GENEC model grids \citep{Ekstrom2012GridsZ=0.014,Georgy2013Grids0.002,Groh2019Grids0.0004} to zero metallicity. From this grid we select 16 models of initial masses $\mini$\,=\,9,\,12,\,15,\,20,\,30,\,40,\,60,\,85,\,120$\,\msun$, both non-rotating and rotating with initial velocity $\vini=0.4 \, \vcrit$, where  $\vcrit \!\!=\!\! \sqrt{\frac{2}{3}\frac{GM}{R_{\rm pol,crit}}}$ is the break-up velocity at critical rotation and $R_{\rm pol,crit}$ is the polar radius at $\vcrit$. We assume no magnetic field, and due to the zero-metallicity nature of the models we assume no mass loss unless the critical rotation limit is reached. Convective zones are determined using the Schwarzschild criterion, and for the main sequence (MS) and the He-burning phase the convective core is extended with an overshooting parameter $d_{\rm over}/H_P=0.1$, where $d_{\rm over}$ is the distance of overshooting beyond the Schwarzchild boundary and $H_P$ is the pressure scale-height at the edge of the core. For further information on model ingredients we refer the reader to the grid paper \citet{Murphy2021}. We also compute new models with different assumptions about convective mixing to complement our Pop~III grid. These new models are non-rotating models of initial masses $\mini$\,=\,9,\,15,\,20,\,30,\,40,\,60,\,85,\,120$\,\msun$ with a higher overshooting parameter of $d_{\rm over}/H_P=0.3$.

In addition, we compute new stellar structure models using \textsc{ \small SNAPSHOT} \citep{Farrell2020_snapshot} to study the impact of other mixing properties. For this we take the abundance profile of a non-rotating 20$\,\msun$ model halfway through the MS and modify the abundance profile to mimic stronger or weaker mixing processes. This involves creating a smoother or steeper H profile above the H burning core. Once the abundance profile has been modified the star is allowed to relax to hydrostatic and thermal equilibrium. For more details on this method we refer the reader to \citet{Farrell2020_snapshot}.

To calculate the total number of ionizing photons produced by these models during their lifetime, we must first calculate the radiative flux produced at each timestep. We approximate the radiative flux at each timestep using a blackbody emitter with the luminosity and effective temperature predicted by the Geneva stellar evolution model, which is obtained using a grey-atmosphere approximation. This caveat of using blackbodies should be taken into account when interpreting our results. The radiative flux of stars is well known to deviate from blackbodies, in particular at lower temperatures and/or when opacity effects are important. Previous work from our group investigated the time evolution of the stellar flux and spectral lines for a 60$\,\msun$ star at solar metallicity \citep{Groh2014}. In that work we merged the Geneva stellar evolution code with radiative transfer calculations using the CMFGEN code \citep{HillierMiller1998} and predicted the flux in different bands, including the emission of ionizing photons. These are computationally-demanding computations that we will defer for future work. For Pop~III stars the main effect of not using stellar atmospheres will be seen in the emission of He~II (and to a lesser degree He~I) ionizing photons as we discuss later in this paper. The number of H ionizing photons produced by Pop~III stars is only weakly affected by using blackbodies ($\lesssim\!0.2$ dex), since H is fully ionized in the atmospheres of all the models that we analyze here. We discuss this point further in \Cref{subsec:Sch02}, where we compare our results to those from \cite{Schaerer2002}. The qualitative results of our paper should not be substantially affected by using blackbodies. 

Once the radiative flux $F_\lambda$ has been computed under the assumptions above, the ionizing photon production rate is computed by integrating below the threshold wavelengths $\lambda_i$ for ionizing photons of HI, HeI and HeII. These wavelengths are 912, 504, and 228$\angstrom$ for HI, HeI, and HeII respectively. To calculate the ionizing photon production rate, $Q_{\rm i}$, in photons s$^{-1}$ \citep{TumlinsonShull2000,Schaerer2002}, we use

\vspace{-0.3cm}
\begin{equation}
\label{eq:Q}
    Q_{\rm i}=\frac{4\pi}{hc} R_\star^2 \int_{0}^{\lambda_i} \lambda\,F_\lambda\,d\lambda
\end{equation}

where $R_\star$ is the stellar radius. The number of ionizing photons produced by a model of a given initial mass during their full lifetime ($\Ni$) is then computed by integrating $\Qi$ across its stellar lifetime.

To calculate the number of stars formed at various initial masses in a population, we use the initial mass function (IMF) of the form $\xi(M) = \xi_0 M^{-\alpha}$, where $\alpha$ is the slope of the IMF, and $\xi_0$ is a factor that depends on the total mass of the population. For most of this work we assume a total mass of the population of $10^6\,\msun$. Through varying the slope of the IMF we can produce different populations where different initial masses will dominate ionizing photon production. The slope used varies from $\alpha=-1$ to $\alpha=2.35$ to cover a range of potential IMFs from top-heavy ($\alpha\!<\!0$ \citealt{Greif2011}; $\alpha\!<\!2$ \citealt{Bromm2012}) to the Salpeter IMF ($\alpha\!=\!2.35$; \citealt{Salpeter1955}) which is widely accepted as the IMF for solar-metallicity stars. While we explore the ionizing photon production of stellar populations of these various IMF slopes, we will focus on $\alpha=0.17$, which is found in \cite{StacyBromm2013} (see \Cref{sec:intro}), to represent the IMF for a Pop~III stellar population. To determine the total number of ionizing photons produced by the population ($N_{\rm pop}$)  we use the following equation, 

\begin{equation}
\label{eq:Npop}
    N_{\rm pop}= \int_{M_{\rm min}}^{M_{\rm max}} \int_{0}^{t} Q_{\rm i}\,\xi(M)\,dt\,dM
\end{equation}

For most of the paper we will assume a minimum mass ($\mmin$) of 9$\,\msun$, and a maximum mass ($\mmax$) of 120$\,\msun$. In \Cref{subsec:results:Mmin} we test the effect of $\mmin$ on the ionizing photons produced. We do this by using our intermediate mass models from \cite{Murphy2021} of $\mini$\,=\,1.7,\,2,\,2.5,\,3,\,4,\,5,\,7$\,\msun$. We also test the effect of $\mmax$ in \Cref{subsec:results:Mmax} by using newly computed zero-metallicity models of initial masses $\mini$\,=\,180,\,250,\,300,\,500$\,\msun$ (Martinet et al., in prep). These very massive star models are non-rotating, and unlike our Pop~III grid models use opacity tables of \protect\cite{GrevesseNoels1993} rather than \protect\cite{Asplund2005}, and use the Ledoux criterion for convective boundaries with convective overshooting of $\alpha_{ov}=0.2$. We note that these massive stars have very large convective cores and thus the differences in the implementation of the physics with respect to the models in \cite{Murphy2021} are not very relevant here. The very massive star models are also run only to the end MS, so we assume that the total photons produced by the end of the MS account for 90\% of the total photons produced by the end of the evolution. This assumption gives an upper estimate to the ionizing photons produced given that models tend to have lower effective temperatures in post MS phases.

The time interval for integration in \cref{eq:Npop}, $t$, refers to the age of the population. Since stellar lifetime increases with decreasing initial mass $\mini$, $t$ must be at least the lifetime of the smallest initial mass model in order for all of the stars in the population to have produced their total ionizing photons. Therefore, if considering a population without rotation where $\mmin=9\,\msun$, the time interval must be $t$\,=\,20\,Myr, the lifetime of the non-rotating 9\,$\msun$ model, for the population to have produced its total ionizing photons after a single starburst, i.e. $N_{\rm pop}= \int_{M_{min}}^{M_{max}} \Ni\,\xi(M)\,dM$. However, we can also vary the value of $t$ to study the evolution of the ionizing photons produced by the population, as we will discuss in \Cref{subsec:results:IMFevol}.


\section{Results}\label{sec:results}

\subsection{Analytical prescription of total ionizing photons produced by Pop~III stars}\label{subsec:results:analyticalfits}

\begin{figure}
    \centering
    \includegraphics[width=\linewidth]{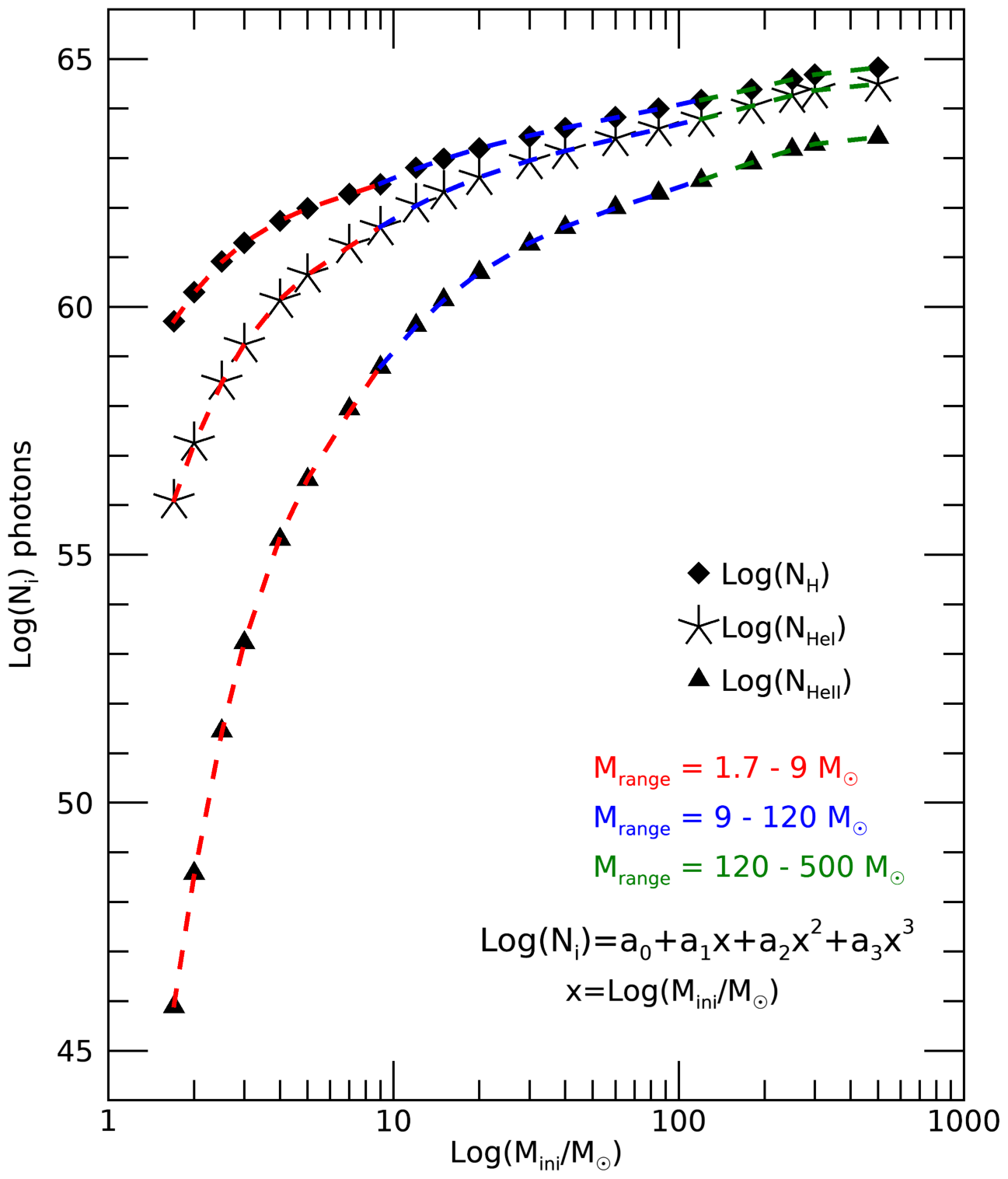}
    \caption{Total ionizing photons produced, $\log(\Ni)$,  by individual non-rotating models in the mass range $1.7\,\msun\leq\mini\leq500\,\msun$. The different symbols correspond to photons capable of ionizing H (diamonds), He~I (asterisks), and He~II (triangles). Also indicated are the mass ranges corresponding to the different fits, and the equation of the cubic polynomial fit. The coefficients of each of the six fits are presented in \Cref{fits_table}.}
    \label{fig:analyticalfits}
\end{figure}

We first present our analytical fits of the total ionizing photons produced by non-rotating models in the full mass range $1.7\,\msun\!\leq\!\mini\!\!\leq500\,\msun$. 
These fits will be useful for future studies and allow for convenient calculation of primordial ionizing photon production in hydrodynamical simulations. In \Cref{fig:analyticalfits} we plot the total ionizing photons produced, $\log(\NH)$, $\log(\NHeI)$ and $\log(\NHeII)$, versus the initial mass, $\log(\mini/\msun)$, along with their least-squares polynomial fits for the mass ranges 1.7-9$\,\msun$, 9-120$\,\msun$, and 120-500$\,\msun$. These fits are described by $\log(\Ni)=a_0+a_1x+a_2x^2+a_3x^3$, where $x=\log(\mini/\msun)$, and the coefficients for each fit are presented in \Cref{fits_table}. We note that unlike \cite{Schaerer2002} these values are based on a blackbody approximation. For zero-metallicity stars the main effect of not using stellar atmospheres will be seen in the emission of He~II (and to a lesser degree He~I) ionizing photons. H ionizing photons are not expected to be significantly impacted since H is fully ionized in the atmospheres of our models. We investigate the impact of this blackbody approximation in \Cref{subsec:Sch02}. In the following sections we study how rotation and convection impact the ionizing photon production. These analytical fits, in conjunction with our predictions for variations with evolutionary parameters, can be used to inform future studies on Pop~III ionizing radiation.

\begin{table}
 \centering   
 \caption{Coefficients of least squares polynomial fits of total ionizing photons produced, $\log(\Ni)$, by non-rotating models in the full mass range $1.7\,\msun\leq\mini\leq500\,\msun$, of the form $\log(\Ni)=a_0+a_1x+a_2x^2+a_3x^3$, where $x=\log(\mini/\msun)$. The mass range is divided into three ranges to cover the intermediate mass models 1.7-9$\,\msun$ \citep{Murphy2021}, the massive models 9-120$\,\msun$ \citep{Murphy2021}, and the very massive models 120-500$\,\msun$ (Martinet et al., in prep). These fits are illustrated in \Cref{fig:analyticalfits}.}
\begin{tabular}{|c|r|c|r|r|r|}
    \hline
    \noalign{\smallskip}
     Quantity  & $\mini$ \hspace{0.2cm} & $a_0$ & $a_1$ \hspace{0.1cm} & $a_2$ \hspace{0.1cm} & $a_3$ \hspace{0.1cm} \\
    \noalign{\smallskip}
    \hline    
    \noalign{\smallskip}
    $\log(\NH)$  &  1.7-9$\,\msun$  &     56.97  &     15.26  &    -16.02  &      6.36  \\
    $\log(\NH)$  &  9-120$\,\msun$  &     57.81  &      7.89  &     -3.82  &      0.72  \\
    $\log(\NH)$  &  120-500$\,\msun$  &     96.12  &    -44.18  &     19.84  &     -2.88  \\
    $\log(\NHeI)$  &   1.7-9$\,\msun$  &     50.76  &     29.54  &    -30.19  &     11.68  \\
    $\log(\NHeI)$  &   9-120$\,\msun$  &     54.53  &     12.11  &     -5.95  &      1.09  \\
    $\log(\NHeI)$  &   120-500$\,\msun$  &     92.51  &    -41.09  &     18.96  &     -2.81  \\
    $\log(\NHeII)$  &    1.7-9$\,\msun$  &     33.55  &     68.19  &    -69.13  &     26.64  \\
    $\log(\NHeII)$  &    9-120$\,\msun$  &     44.96  &     23.64  &    -11.52  &      2.03  \\
    $\log(\NHeII)$  &    120-500$\,\msun$  &     86.11  &    -36.27  &     17.61  &     -2.70  \\

    \noalign{\smallskip}
    \hline    
    \noalign{\smallskip}
    \end{tabular}
    \label{fits_table}

 \end{table}

\subsection{Rotation: Impact on ionizing photon production rate}\label{subsec:results:rotationQi}

\begin{figure*}
    \centering
    \includegraphics[width=\linewidth]{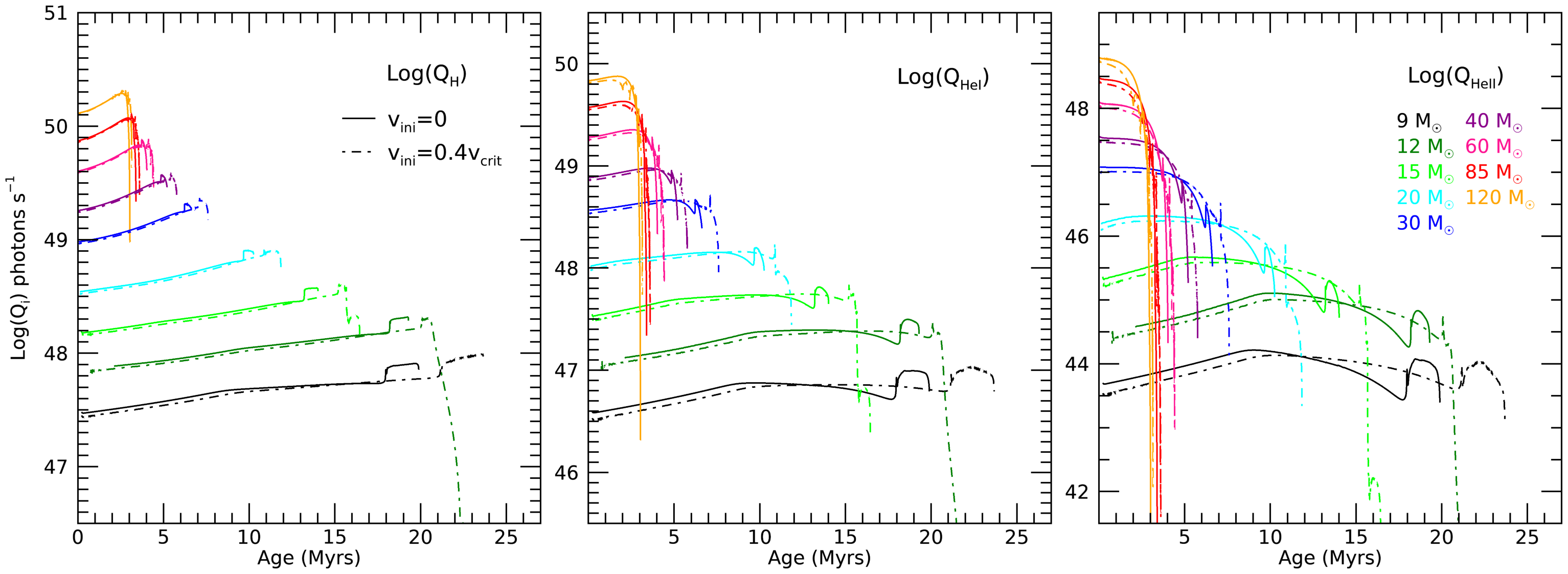}
    \caption{Evolution of the ionizing photon production rate ($\Qi$) for models in the mass range $9\,\msun\! \leq\! \mini\! \leq\! 120\,\msun$. Three ionizing photons species are shown, H (left panel), He~I (middle panel), He~II (right panel). Solid lines show non-rotating models, dashed lines show rotating models, and colours indicate different initial masses as shown in legend.}
    \label{fig:Qphots_evol}
\end{figure*}

We now look at how the ionizing photon production rate, $\Qi$, varies throughout the evolution of our Pop III models from \cite{Murphy2021}. \Cref{fig:Qphots_evol} shows this evolution for photons capable of ionizing H, He~I and He~II. Higher initial mass models can be distinguished by their shorter stellar lifetimes. We see that the higher the initial mass the larger the ionizing photon production rate. This is not surprising given that more massive models are typically hotter and more luminous than models of lower initial mass. \Cref{fig:Qphots_evol} shows that there are two competing effects in determining how many ionizing photons will be produced by a model during its lifetime: the ionizing photon production rate, which depends on the surface properties; and the stellar lifetime, which limits the time available for producing ionizing photons. Models including rotation are indicated by dashed lines and are noticeable for their longer lifetimes compared to non-rotating models (solid lines). We would expect that this will then increase the total number of ionizing photons produced by rotators, but this is not clear given that the ionizing photon production rate varies due to differences in surface properties with rotation.

\begin{figure*}
    \centering
    \includegraphics[width=\linewidth]{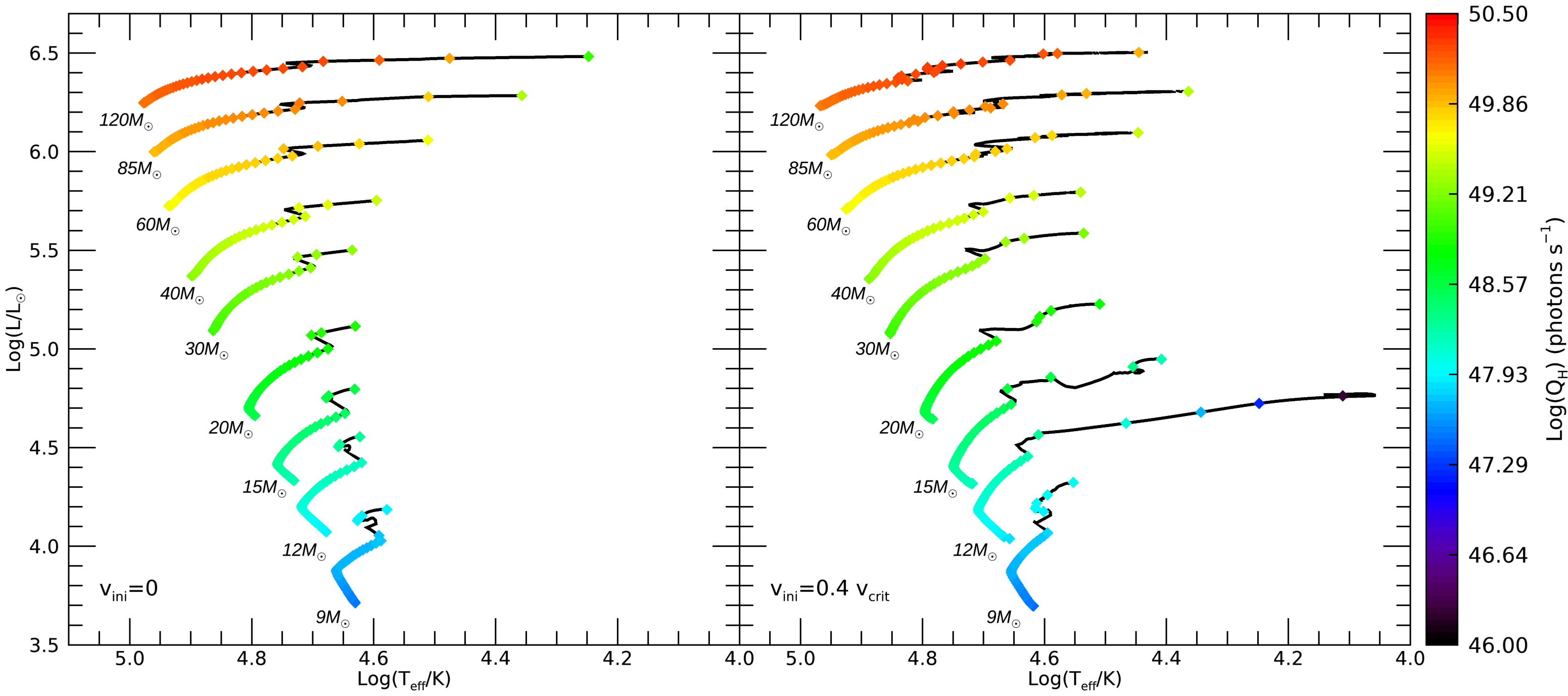}
    \caption{\textit{Left:} Evolutionary tracks of non-rotating models from \protect\cite{Murphy2021} in the mass range $9\,\msun\!\leq\!\mini\!\leq\!120\,\msun$. Overplotted are the values for the H ionizing photon production rate, $\log(\QH$), as indicated by the colour bar on the right-hand side. \textit{Right:} Same as left panel but for models rotating with initial velocity $\vini=0.4\,\vcrit$.}
    \label{fig:HRD}
\end{figure*}

To further illustrate how the ionizing photon production rate varies with different initial masses we plot the production rate of H ionizing photons, $\log(\QH$), over the evolutionary tracks of the models in \Cref{fig:HRD}. The values of $\log(\QH$) are taken at equal intervals of age, so this also has the advantage of visualising the time evolution across the Hertzsprung-Russell diagram, and where each model spends the majority of its time producing ionizing photons. Since these stars spend approximately 90\% of their lifetimes on the MS, it is not surprising that they produce the majority of their ionizing photons there. From this figure we can see how the H ionizing photon production rate varies in different regions of luminosity and effective temperature, and visualise why models of higher initial mass have higher ionizing photon production rates. We can also observe how rotational effects on the surface properties result in changes to the ionizing photon production rate. For example, the 12$\,\msun$ model with rotation experiences a decrease in surface temperature ($\teff$) at late evolutionary stages which results in a large decrease in $\QH$.

\begin{figure*}
    \centering
    \includegraphics[width=\linewidth]{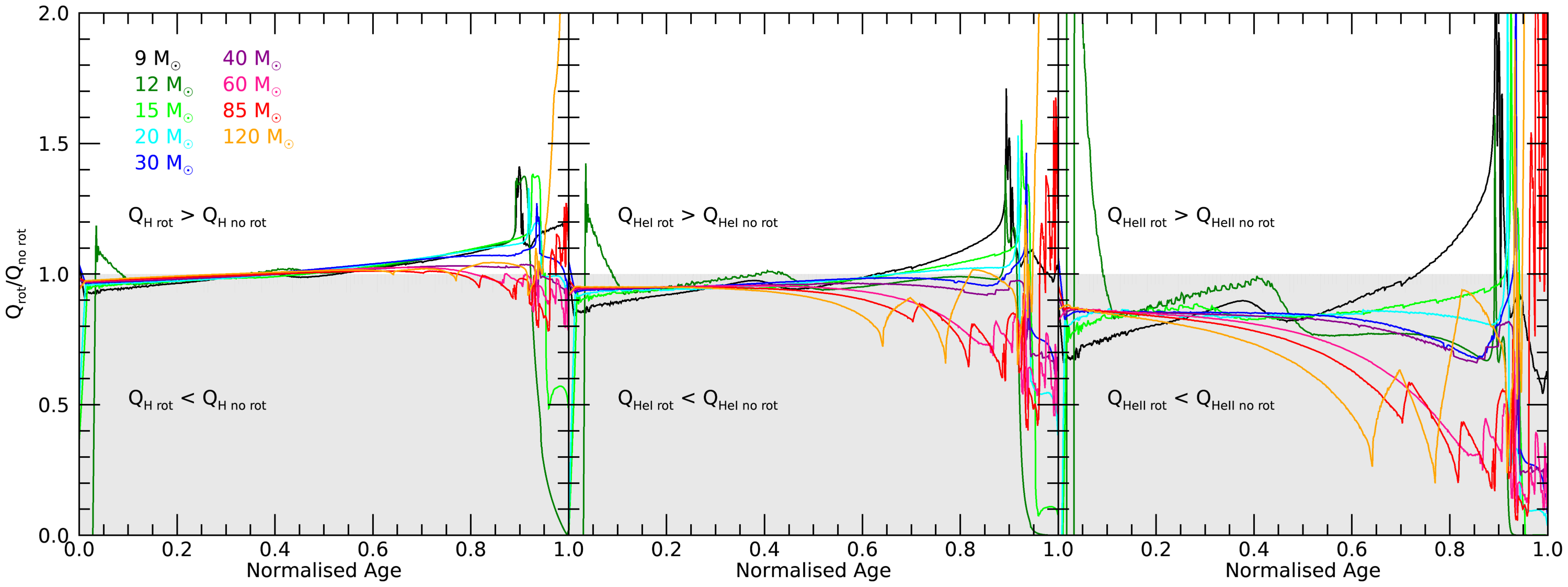}
    \caption{\textit{Left:} Evolution of ratio of H photon production rate, $\QH$ (\cref{eq:Q}), by models with rotation to models without rotation. Colours represent different initial masses, $\mini$. The white region indicates where $\QH$ is higher with rotation, conversely the grey region indicates where non-rotating models have higher $\QH$. \textit{Middle:} Same as left panel but for He~I photons. \textit{Right:} Same as left panel but for He~II photons.}
    \label{fig:Qratio}
\end{figure*}

To better understand how the ionizing photon production rate, $\Qi$, varies with rotation we calculate the ratio of $\Qi$ for models with rotation to those without, $\frac{\Qrot}{\Qnorot}$, for each of the three ionizing photon species, H, He~I, and He~II. In order to compare their time evolutions properly, this is done for the normalised age of each model. These ratios are shown in \Cref{fig:Qratio} for each initial mass in the range  $9\,\msun\! \leq\! \mini\! \leq\! 120\,\msun$. It can be seen that rotation impacts the production rate of each species differently. In the case of H photons (left panel \Cref{fig:Qratio}), while rotating models start their lives producing less H photons, the ratio of $\Qrot$ to $\Qnorot$ increases steadily through most of the lifetime, although there is some divergence in late stages for the more massive models. Focusing now on He~I photons (middle panel \Cref{fig:Qratio}) we see an earlier divergence in the trend for varying initial masses. From about half-way through the lives of the models, less massive models see an increase of $\Qrot$ to $\Qnorot$ while more massive models see a decrease in this ratio. We also note that, with the exception of the late stages of models with $\mini$=9-20$\,\msun$, more He~I photons are produced by non-rotating models. Finally we look at the effect of rotation on the production of He~II photons. The changes in behaviour moving from H to He~I photons seem amplified here. That is to say that the ratio of $\Qrot$ to $\Qnorot$ has decreased even further, with more massive rotating models producing as little as half the He~II photons as their non-rotating counterparts for a significant fraction of the lifetime. These evolving trends for different photon species call into question how these ionizing photon production rates depend on the surface properties.

\begin{figure}
    \centering
    \includegraphics[width=\linewidth]{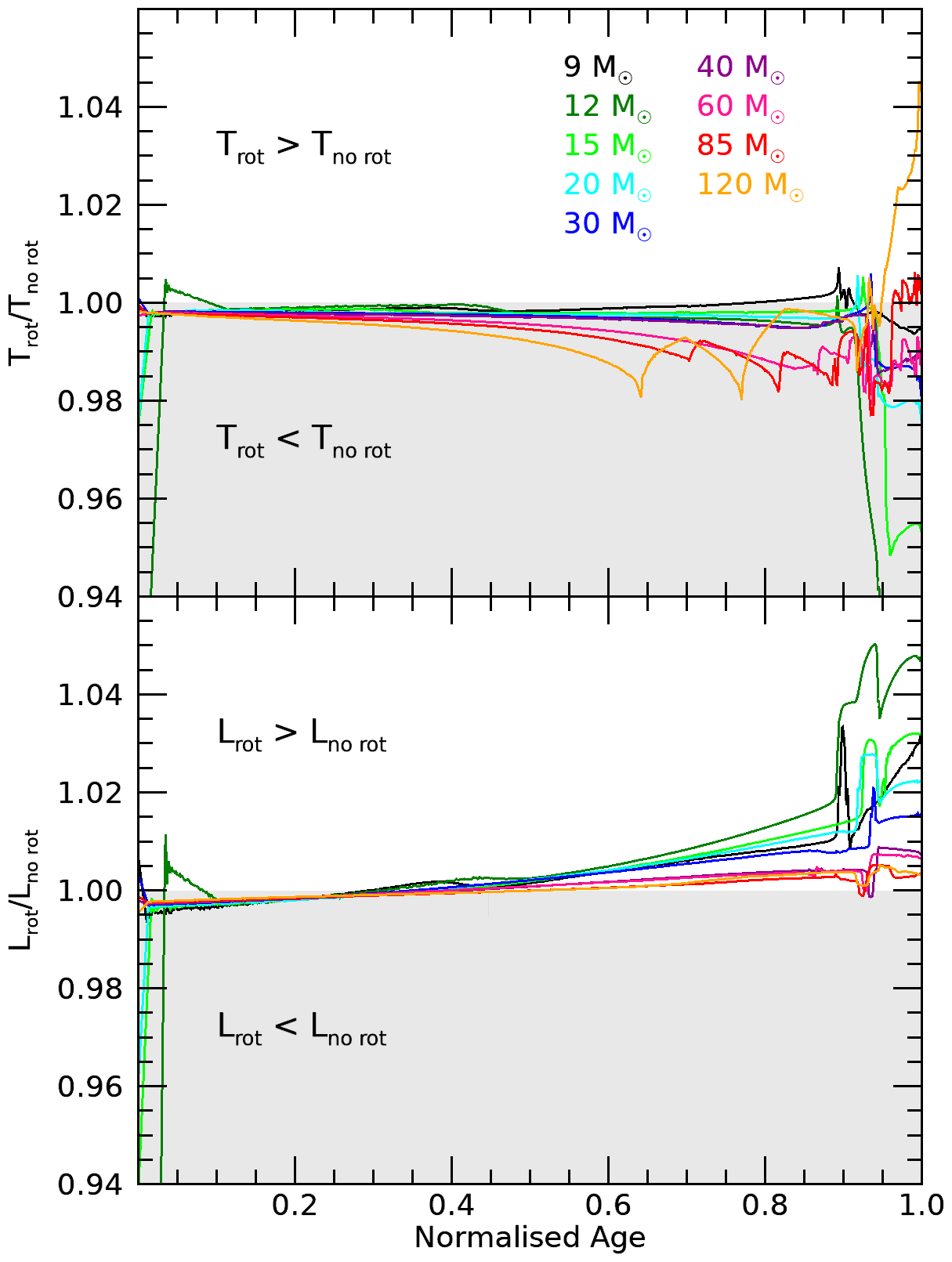}
    \caption{\textit{Upper:} Evolution of ratio of effective temperature ($\teff$) by models with rotation to models without rotation. Colours represent different initial masses, $\mini$. The white region indicates where $\teff$ is higher with rotation, conversely the grey region indicates where non-rotating models have higher $\teff$. \textit{Lower:} Same as upper panel but for luminosity.}
    \label{fig:TLratio}
\end{figure}

To understand the trends seen in \Cref{fig:Qratio} and what leads to differences from species to species, we look at the impact of rotation on surface properties throughout the normalised lifetimes. This is shown in \Cref{fig:TLratio}, again using ratios to disentangle where rotating models are hotter or more luminous than non-rotating models, and vice versa. We see separate trends here for these two surface properties. On the one hand, the effective temperature ($\teff$) of rotating models tends to decrease over the evolution relative to non-rotating models, a trend which is more strongly seen as initial mass increases. This decrease in $\teff$ for more massive models, $\mini\!=\!60,85,120\,\msun$, corresponds to their approach towards critical rotation (see Figure 6, \citealt{Murphy2021}). On the other hand, the luminosity of rotating models tends to increase over the evolution relative to non-rotating models, and in contrast to the trend for temperature this is more strongly seen for less massive models. This explains two things from \Cref{fig:Qratio}. Firstly, it explains the dichotomy of trends with initial mass at late stages where the impact of rotation appears to diverge, and secondly, it illustrates which of the surface properties dominates the production rate of each ionizing photon species. We can now observe that H photons are dominated by the luminosity of the model since the trends seen in the left panel of \Cref{fig:Qratio} most closely resemble the trends seen in the lower panel of \Cref{fig:TLratio}. Similarly we deduce that He~II photons are dominated by effective temperature, while the dependencies of He~I photons on the surface properties lie somewhere in between. This is a reflection of how each species of photon is determined in the first place, and the sensitivity of the stellar radiative flux to the effective temperature in different wavelength domains. As discussed in \Cref{sec:methods}, we use a blackbody approximation to obtain the radiative flux and subsequently integrate below the different threshold wavelengths to determine the ionizing photon production rate for each species. The lower the wavelength, the larger the effect of changing the effective temperature on the number of photons at that wavelength. This phenomenon would be similar had we used detailed atmospheres.

\subsection{Rotation: Impact on total ionizing photons produced}\label{subsec:results:rotationNi}

\begin{figure*}
    \centering
    \includegraphics[width=0.85\linewidth]{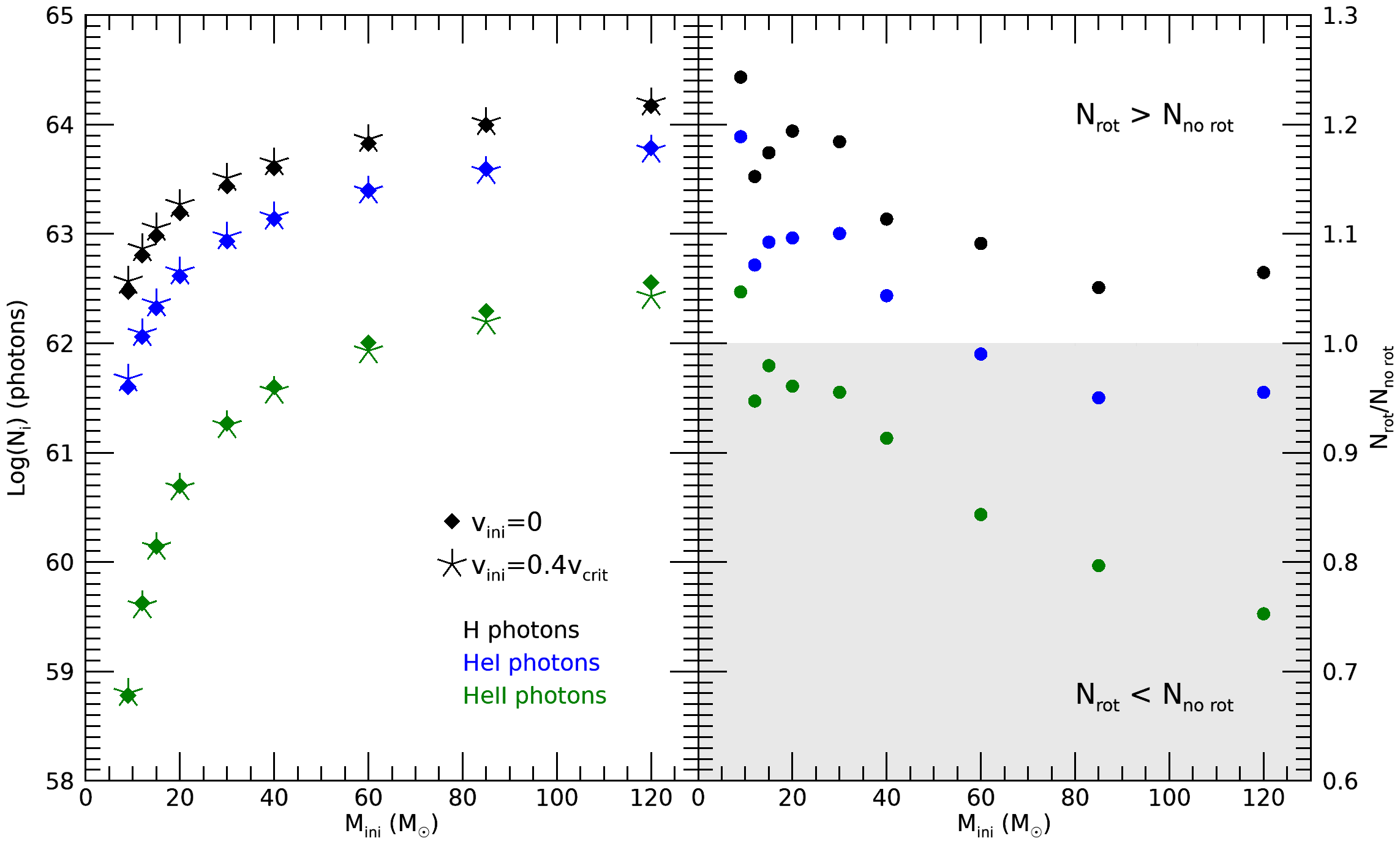}
    \caption{\textit{Left:} Total ionizing photons produced from the zero age main sequence (ZAMS) to end He-burning as a function of $\mini$. Models with (without) rotation are indicated by a diamond (asterisk) symbol. Colours show the different ionizing photon species. \textit{Right:} Ratio of total ionizing photons produced by models with rotation to models without rotation as a function of $\mini$. Similarly to \Cref{fig:Qratio}, the grey region shows where rotating models produced less ionizing photons from the ZAMS to the end of He-burning, and the white region shows where more ionizing photons were produced by rotators.}
    \label{fig:Nphot_perstar}
\end{figure*}

As has been identified in \Cref{subsec:results:rotationQi}, while $\Qi$ increases with initial mass, more massive models also have shorter lifetimes, and therefore the outcome for total ionizing photons produced by a star of a given initial mass ($\Ni$) depends on the combination of these two effects. In \Cref{fig:Nphot_perstar} we present the results for total ionizing photons produced for each initial mass model, with and without rotation. These values are also presented in \Cref{initial_mass_table}. From this figure it is clear that the total number of ionizing photons increases for increasing initial mass, despite the decrease in stellar lifetimes. This is true for all three species of ionizing photons. The trend with initial mass is apparent from the left panel of this figure, however, the trend with rotation is more difficult to observe. To clarify this we present in the right panel of \Cref{fig:Nphot_perstar} the ratio of total ionizing photons produced by rotating and non-rotating models of a given initial mass.

The complexity of the trends in this figure reflects the complexities of the impact of rotation on the surface properties. The effect of rotation is not the same for each initial mass, and each species is affected differently by rotation, following on from what we have discussed in \Cref{subsec:results:rotationQi}. For each ionizing photon species we see a decrease in the ratio $N_\mathrm{rot}/N_\mathrm{no rot}$ as we move to higher initial masses. For the 9$\,\msun$ model, we see that rotation increases the total photons produced for all three species, with the rotating model producing 25\% more H photons than the non-rotating 9$\,\msun$ model. For all other initial masses rotating models produce less He~II photons than non-rotating models, with the 120$\,\msun$ rotating model producing 25\% less He~II photons than the non-rotating model. This result is important, because it tells us that not only will the total ionizing photons produced change for differing initial masses, but the rotational effects vary also. The impact of rotation on surface properties and stellar evolution of the first stars is discussed in detail in \cite{Murphy2021}. While rotational effects are complex and differ significantly with initial mass, the dominant effects are increasing luminosity, due to larger convective cores, and decreasing surface temperature, due to changes to stellar structure. The outcome of these two competing effects varies with initial mass. We saw this in \Cref{fig:TLratio}, where decreasing surface temperature with rotation was more prominent for higher initial masses, and increasing luminosity with rotation was more prominent for lower initial masses. From \Cref{subsec:results:rotationQi} we found that H photons are dominated by luminosity effects, and He~II photons are dominated by surface temperature effects. This is why in the right panel of \Cref{fig:Nphot_perstar} we see a stronger change to H photons with rotation at lower initial masses, and a stronger change to He~II photons at higher initial masses.
The mass dependency of rotational effects is thus evident in the ionizing photon species most impacted. This effect should be considered in studying the impact of the initial mass function on ionizing photons produced.

\subsection{Convection: Impact on total ionizing photons produced}\label{subsec:results:convection}

\begin{figure*}
    \centering
    \includegraphics[width=0.85\linewidth]{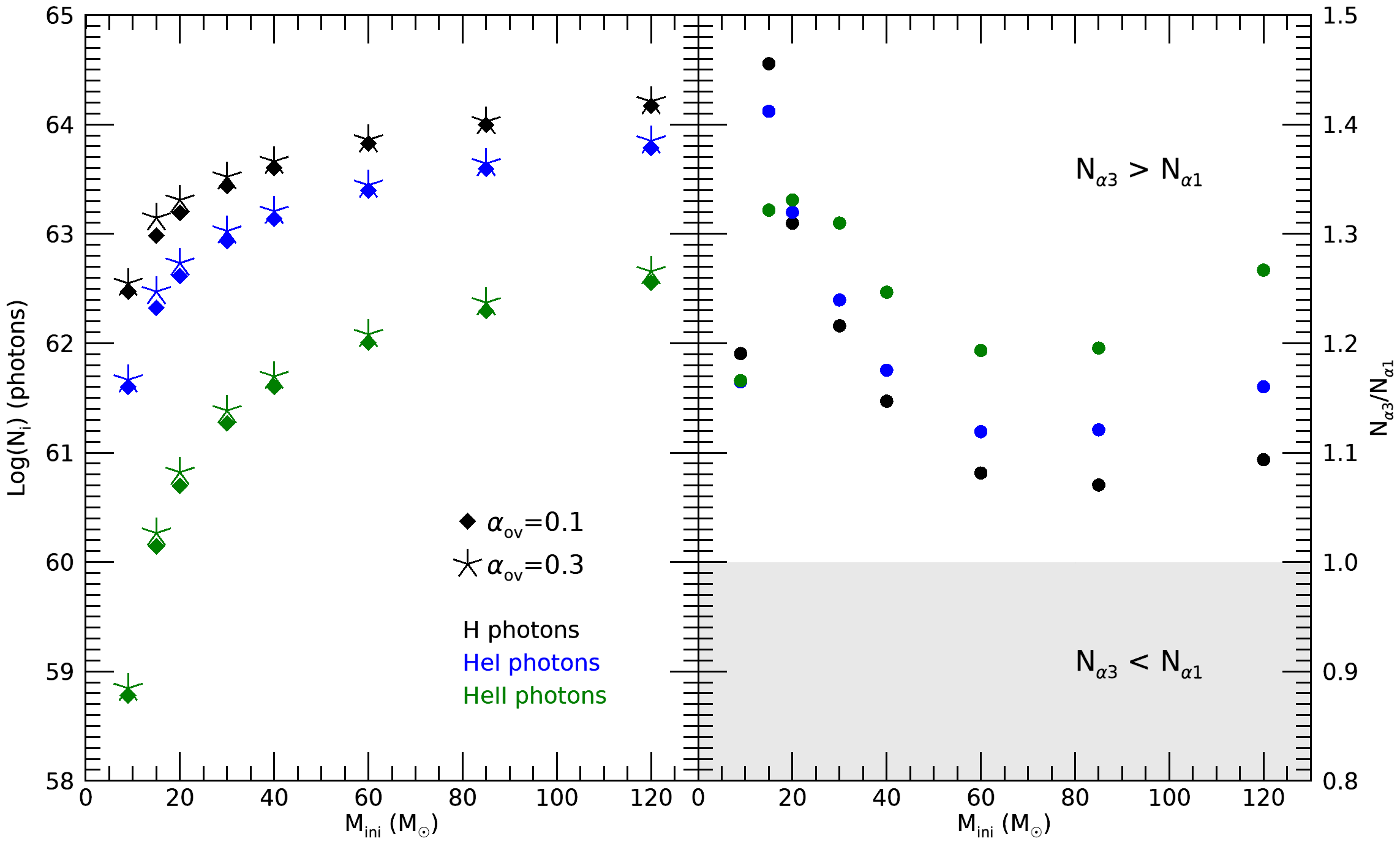}
    \caption{\textit{Left:} Total ionizing photons produced by models with convective overshooting parameter $\alpha_{ov}=0.1$ (diamond symbols) and $\alpha_{ov}=0.3$ (asterisk symbols). \textit{Right:} Ratio of total ionizing photons produced by models with $\alpha_{ov}=0.3$ to models with $\alpha_{ov}=0.1$.}
    \label{fig:Nphot_convection}
\end{figure*}

Similarly to rotation, convective overshooting above the core increases interior mixing and impacts stellar evolution significantly. For consistency with previous Geneva stellar evolution grids \citep{Ekstrom2012GridsZ=0.014,Georgy2013Grids0.002,Groh2019Grids0.0004} the model grid used here \citep{Murphy2021} takes a value of $\alpha_{ov}=0.1$ for the overshooting parameter. However, in recent research it has been predicted that the overshooting parameter could be higher for massive stars, with $\alpha_{ov}=0.3-0.5$ more closely matching observations of massive MS stars \citep{Castro2014overshoot,Schootemeijer2019,HigginsVink2019,Martinet2021}. We therefore want to investigate the impact that increased convective overshooting will have on ionizing photon production. To do this we take additional Geneva stellar evolution models of Pop III stars with consistent physical ingredients to our \citealt{Murphy2021} non-rotating grid, barring the overshooting parameter which in this new set of models is $\alpha_{ov}=0.3$. By comparing these models we can discern the effect that increased convective overshooting has on the surface properties, and subsequently the ionizing photon production. The results of this investigation are presented in \Cref{fig:Nphot_convection}, with values given in \Cref{initial_mass_table}. We show the total ionizing photons produced by each model in the left panel, while the right panel shows the ratio of photons produced by the models with higher overshooting, to those with the lower overshooting parameter in the original model grid. We find that for all ionizing photon species, at all initial masses considered here, that increased convective overshooting increases the total ionizing photons produced. This increase varies for different initial masses, but generally speaking we find an increase of approximately 20\% to ionizing photons produced. This result is a reflection of the increased luminosity and surface temperature of the models with higher convective overshooting, but mainly results from the increased lifetime of models with higher overshooting. This is evident from the variations in the percentage increase of ionizing photons for different initial masses. The most notable increase in ionizing photons produced is that of the 15$\,\msun$ model, which experiences the largest increase in MS lifetime with higher overshooting for this initial mass range. Similarly to what we discussed regarding \Cref{fig:Nphot_perstar}, the impact on different ionizing photon species tells us the dominant effect on surface properties for different initial masses. Changes to surface temperature impact He II photons more strongly, while changes to luminosity predominantly impact H photons. Despite these variations for different stellar masses, higher overshooting increases ionizing photon production for all initial masses considered here, which suggests that we can scale the contribution of Pop III stars to ionization with the overshooting parameter considered. 

\subsection{Initial Mass Function: Impact on ionizing photons produced by stellar populations}\label{subsec:results:IMF}

\begin{figure*}
    \centering
    \includegraphics[width=\linewidth]{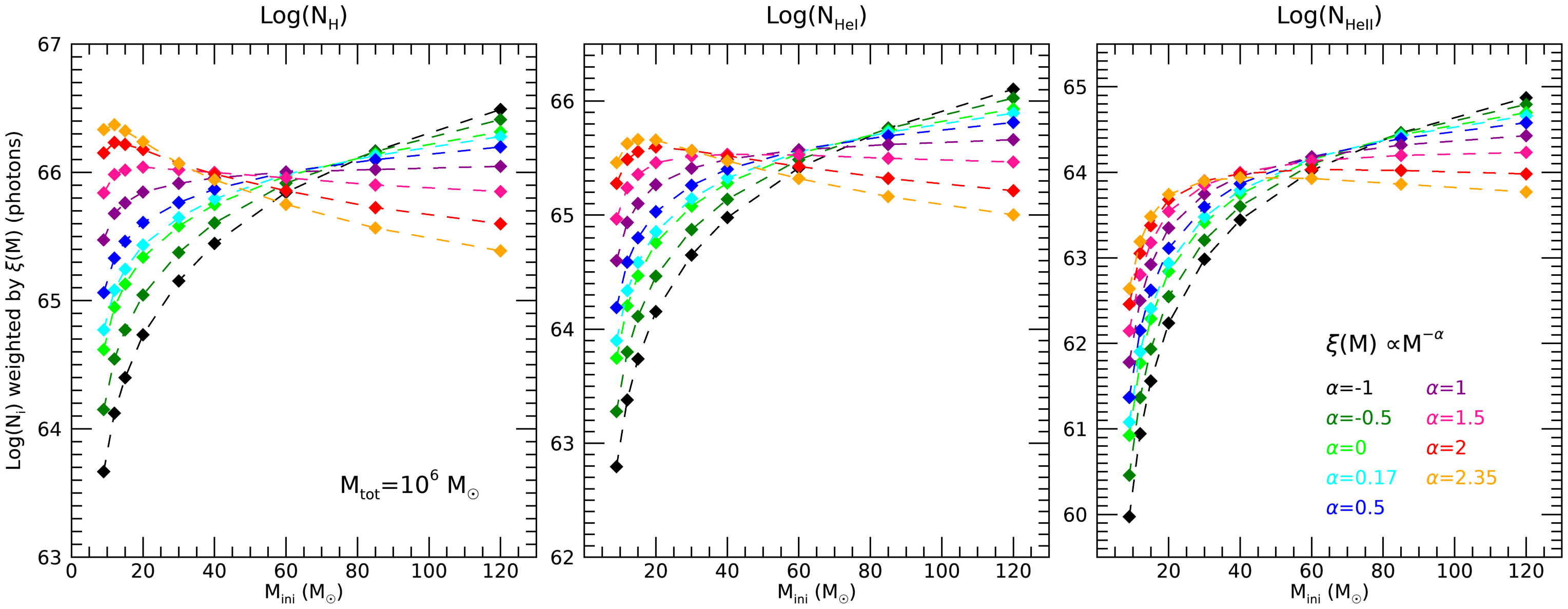}
    \caption{Total ionizing photons produced by non-rotating stars of different initial masses weighted by initial mass functions of the form $\xi(M) \propto M^{-\alpha}$. Each panel represents a different species of ionizing photon, H, He~I, He~II, as indicated by the titles. The total mass of stars in each population is $10^6\,\msun$. Ionizing photons produced at each initial mass vary based on the number of stars produced at that mass by the given IMF. IMF slope values are indicated by the legend.}
    \label{fig:NphotIMF}
\end{figure*}

Now that we understand ionizing photon production for individual stars at zero-metallicity, we turn our attention to the ionizing photon production of populations of these stars. To investigate this we produce initial mass functions (IMFs) of the form $\xi(M)\!\propto\!M^{-\alpha}$ (see \Cref{sec:methods}) and vary the slope, $\alpha$, with a fixed total stellar mass for the population of $\mathrm{M_{tot}}\!=\!10^6\,\msun$. \Cref{fig:NphotIMF} shows the total photons produced by stars of different initial masses weighted by the different IMFs, i.e. $\log(\Ni\xi(M))$. The stars considered here are non-rotating with $\alpha_{ov}\!=\!0.1$. For each IMF, the total photons produced at each initial mass vary depending on the number of stars produced at that initial mass. Furthermore, the total photons produced as depicted here represents the number produced following the lifetime of each model, such that each star has enough time to produce their total number of ionizing photons, $\Ni$ (see \Cref{sec:methods}).
From this figure we find which initial mass model dominates the ionizing photon production for different IMFs. For the steepest IMF, the Salpeter IMF ($\alpha\!=\!2.35$), less massive models dominate photon production, then as you move to lower values of $\alpha$ the more massive models become more important for ionization. This trend holds for each ionizing photon species, however, the more energetic He photons are dominated by higher initial masses than H photons. For the Salpeter IMF ($\alpha\!=\!2.35$), the 12$\,\msun$ stars contribute most to H photon production, the 15$\,\msun$ stars contribute most to He~I photon production, and the production of He~II photons is dominated by 40$\,\msun$ stars. However, for the \cite{StacyBromm2013} (SB13, $\alpha\!=\!0.17$) IMF, the 120$\,\msun$ model dominates H, He~I, and He~II photon production.

\begin{figure*}
    \centering
    \includegraphics[width=0.9\linewidth]{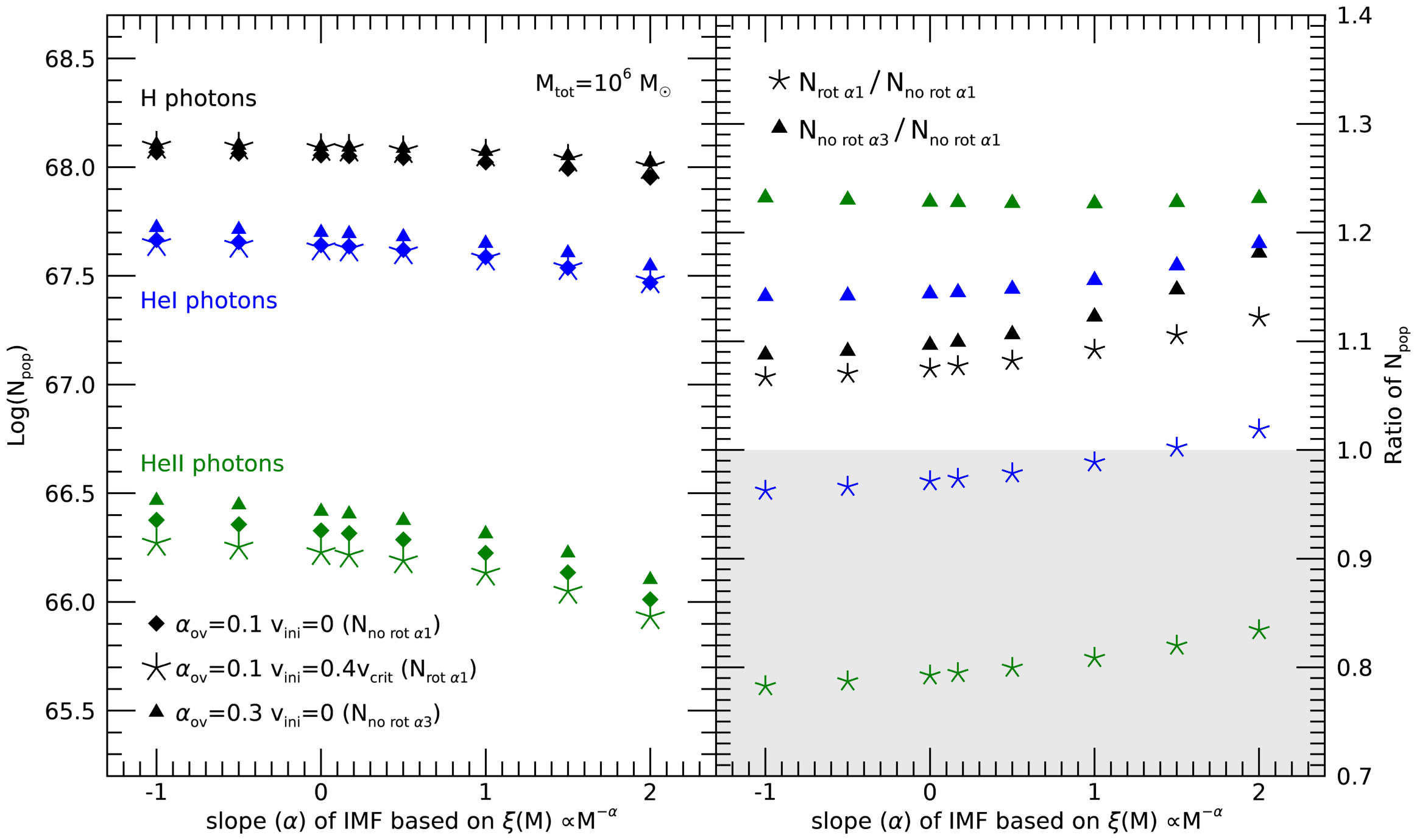}
    \caption{\textit{Left:} Total ionizing photons produced by populations of varying IMF slope $\alpha$. Three model sets are shown. The non-rotating models with overshooting parameter $\alpha_{ov}\!=\!0.1$ (diamond symbols), the rotating models with $\alpha_{ov}\!=\!0.1$ (asterisk symbols), and finally the non-rotating models with higher overshooting $\alpha_{ov}\!=\!0.3$ (triangle symbols). Colours show the different ionizing photon species. The total mass of stars in each population is $10^6\,\msun$. \textit{Right:} Asterisk symbols show the ratio of total ionizing photons produced by models with rotation to models without rotation for each IMF slope, while triangle symbols show the ratio of total ionizing photons produced by models with higher overshooting to models with lower overshooting.}
    \label{fig:Nphot_effectsummary}
\end{figure*}

The impacts of rotation and convection for varying IMF slopes are summarised in \Cref{fig:Nphot_effectsummary}. Similarly to \Cref{fig:Nphot_perstar,fig:Nphot_convection}, the total photons produced for different initial conditions are shown on the left, and ratios of these points are shown on the right. Unlike \Cref{fig:Nphot_perstar,fig:Nphot_convection}, which showed photons produced by individual stars of different initial masses ($\Ni$), this plot shows the total photons produced by stellar populations of different IMF slopes ($\Npop$). We still assume here a population stellar mass of M$_{\mathrm{tot}}\!=\!10^6\,\msun$, and that all stars in the population have produced their total ionizing photons for their full lifetimes, having formed from a single starburst. This figure depicts the importance of the choice of IMF in determining a population's contribution to reionization. As the IMF slope increases, the number of massive stars in the population decreases. Since these more massive stars produce the most ionizing photons, we see a trend of decreasing ionizing photon production with increasing IMF slope. We see this clearly for He~II ionizing photons, however, for H ionizing photons the decrease with increasing IMF slope is much less significant, and may be less than effects due to convection or rotation depending on the IMF slopes being compared. This infers that the H ionizing photon production per solar mass of stars of different initial masses ($\NH/\mini$) is more similar than that of He~II ionizing photon production per solar mass ($\NHeII/\mini$).

The right panel of \Cref{fig:Nphot_effectsummary} shows how effects due to rotation and convection vary with IMF slope. Since the IMF slope determines the dominant initial mass in the population, the rotational effects seen here are a reflection of the trends seen in the right panel of \Cref{fig:Nphot_perstar}, and equivalently the effects of convection seen here are a reflection of the trends seen in \Cref{fig:Nphot_convection}. Lower IMF slopes follow the trend for more massive models, for example a stronger decrease in He~II photons produced by rotating stars. By contrast, as IMF slope increases we start to observe the trend for less massive models, which is a stronger increase in H photons produced by stars with rotation, and stars with higher overshooting. The differences between effects due to rotation and convection in this figure serve as a reminder that rotational effects are more complex than changes in convective core size alone. This reinforces the importance of accurately modelling rotational effects in stellar evolution models, to fully understand the impact that rotation has on stellar structure. 

The main conclusion from \Cref{fig:Nphot_effectsummary} is the relative importance of rotation, convective overshooting, and slope of the initial mass function, in determining the total ionizing photons produced by a population of zero-metallicity stars of a given fixed total mass. Taking the non-rotating $\alpha_{ov}\!=\!0.1$ SB13 population as a reference, the value for total H ionizing photons produced per stellar mass of the population ($\Npop/\mathrm{M_{tot}}$) is 1.13$\times10^{62}$\,photons\,$\msun^{-1}$ (see \Cref{slope_table}). This value increases by 7.7\% when rotation of $\vini\!=\!0.4\,\vcrit$ is included, by 9.9\% with higher overshooting of $\alpha_{ov}\!=\!0.3$, and decreases by 26\% when we assume a Salpeter IMF slope. Therefore, if comparing between the SB13 and Salpeter IMF slopes, then the choice of IMF slope is relatively more impactful than the effect of rotation or convective overshooting for the values considered here.

\subsection{Initial Mass Function: Evolution of ionizing photon production}\label{subsec:results:IMFevol}

\begin{figure}
    \centering
    \includegraphics[width=\linewidth]{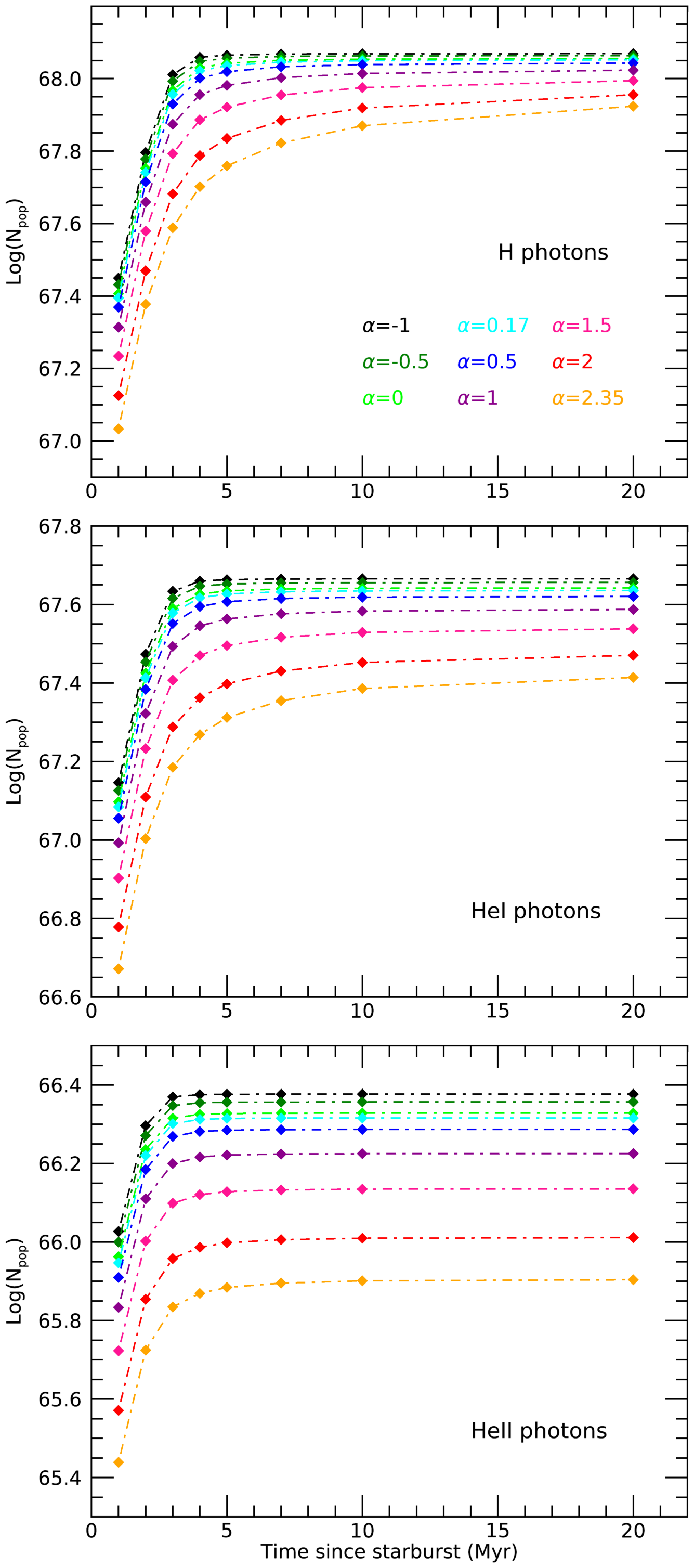}
    \caption{Temporal variation of $\Npop$ (see \cref{eq:Npop}) for different IMF slopes, $\alpha$, indicated by different colours as shown in the legend. The three panels show H (upper), He~I (middle), and He~II (lower) ionizing photons. The time, $t$, considered is the time since the starburst when the non-rotating population with total stellar mass $\mathrm{M_{tot}=10^6\,\msun}$ formed. The mass range here is $9\,\msun\!\leq\!\mini\!\leq\!120\,\msun$, thus at 20$\,$Myr the population has produced its total ionizing photons.}
    \label{fig:Nphot_timeevol}
\end{figure}

Now we investigate the time evolution of the ionizing photons produced by a population of Pop~III stars. The results of this are shown for IMF slopes $-1\leq\alpha\leq2.35$ in \Cref{fig:Nphot_timeevol}, for non-rotating stars with convective overshooting $\alpha_{ov}\!=\!0.1$. At each time, $t$, from 1-20\,Myr after formation, the total ionizing photons produced ($\Npop$) are shown, where the time considered is the time since the starburst when the population with total stellar mass $\mathrm{M_{tot}=10^6\,\msun}$ formed. The mass range here is $9\,\msun\!\leq\!\mini\!\leq\!120\,\msun$, thus at 20$\,$Myr the population has produced its total ionizing photons. Therefore, the rightmost point in each panel of \Cref{fig:Nphot_timeevol} represents the total ionizing photons produced by the population, over the full lifetimes of each star within the population. In the case of the SB13 IMF ($\alpha\!=\!0.17$) we note that the total H ionizing photons produced by the population changes little after 4\,Myr, having produced 93\% of its H ionizing photons by this time. This is because a population with IMF slope $\alpha\!=\!0.17$ is dominated by massive stars with lifetimes of only a few million years. On the other hand, we see a more gradual evolution in total H ionizing photons produced by the Salpeter IMF ($\alpha\!=\!2.35$) population, with 60\% of H ionizing photons produced in the first 4\,Myr, and 94\% of H ionizing photons produced by 10\,Myr. It is clear from \Cref{fig:Nphot_timeevol} that the lower the IMF slope $\alpha$, the faster the ionizing photon production, which is expected given that more massive models have shorter lifetimes. This demonstrates that the chosen IMF of the population could have an important impact on the reionization timescale, with more top-heavy IMFs potentially resulting in an earlier end to the reionization epoch.

\begin{figure}
    \centering
    \includegraphics[width=\linewidth]{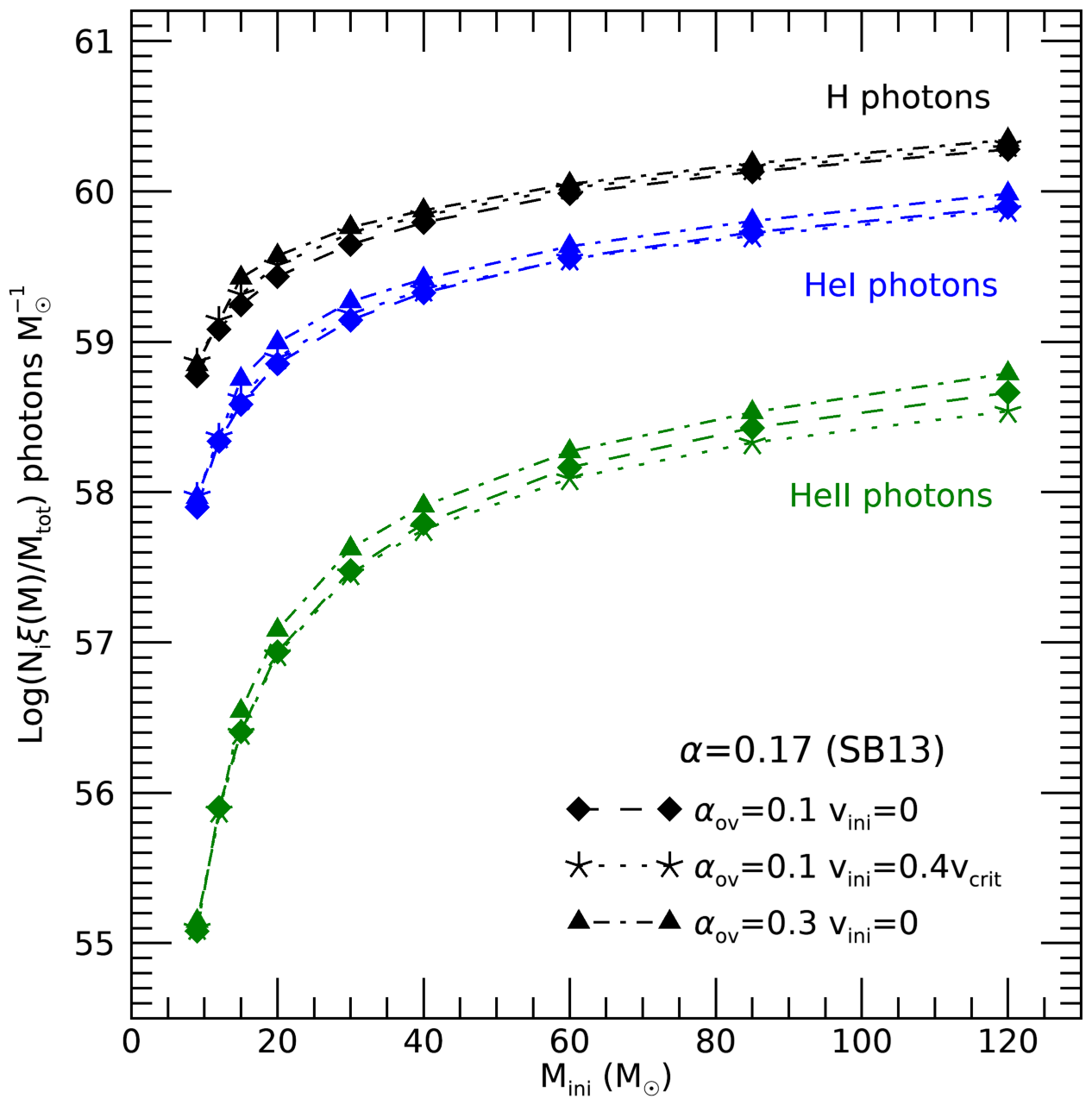}
    \caption{Total number of ionizing photons produced per solar mass of the population, by stars of different initial masses, based on a population of IMF slope $\alpha\!=\!0.17$, i.e. $\log(\Ni\,\xi(M)/\mathrm{M_{tot}})$. Black, blue and green symbols indicate ionizing photon species H, He~I and He~II respectively, as shown in legend. Different symbols and line styles show initial parameters corresponding to rotation and convective overshooting. The values shown here are also presented in \Cref{mass_table_SB13}.}
    \label{fig:Nphot_mass_perMsun}
\end{figure}

\begin{figure}
    \centering
    \includegraphics[width=\linewidth]{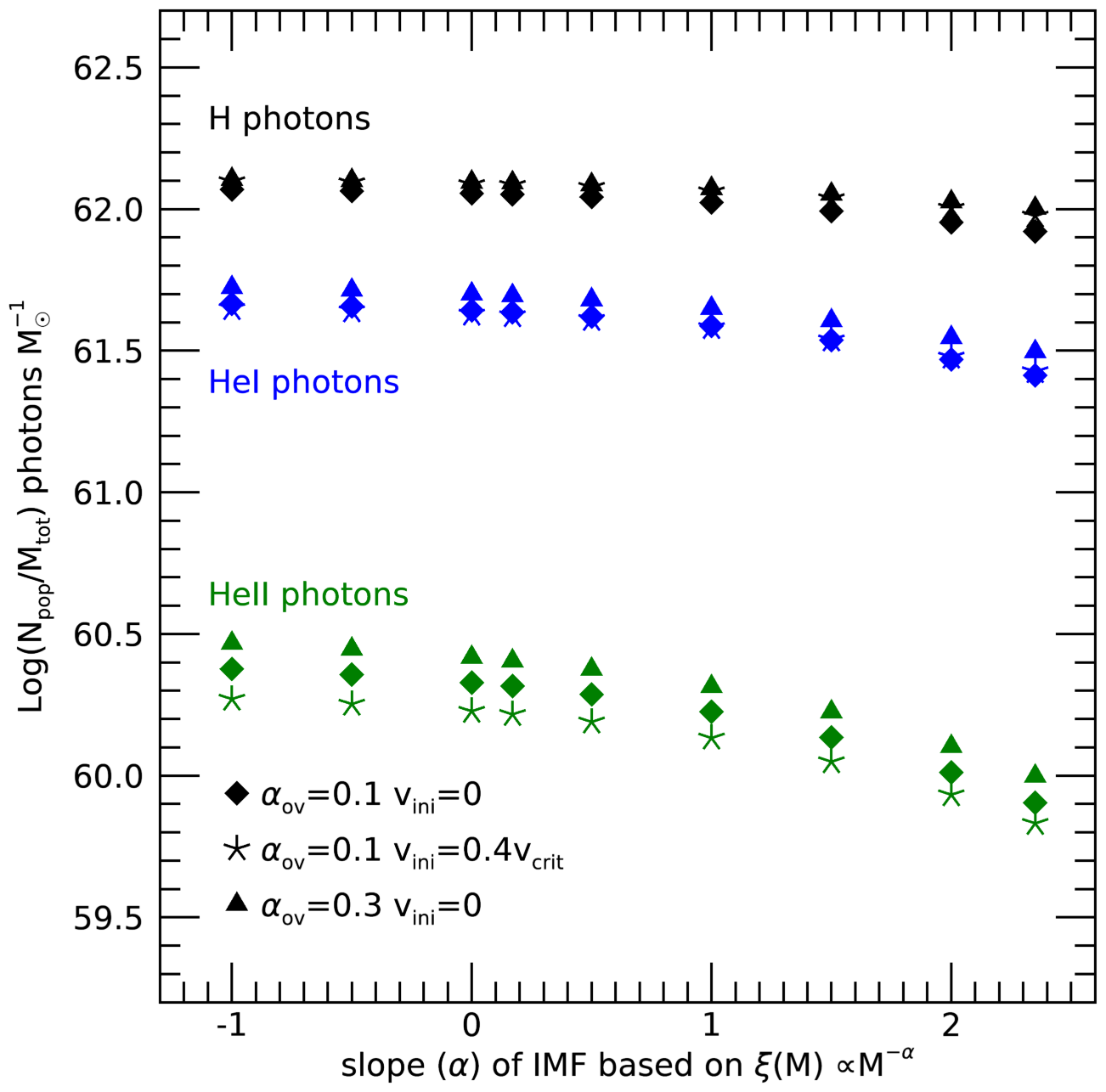}
    \caption{Total number of ionizing photons produced per solar mass of the population, by populations of various IMF slopes. Similarly to \Cref{fig:Nphot_mass_perMsun}, colours indicate different ionizing photon species, and symbols indicate different initial parameters corresponding to rotation and convective overshooting. The values shown here are also presented in \Cref{slope_table}.}
    \label{fig:Nphot_IMF_perMsun}
\end{figure}

We now provide estimates for ionizing photon production in the case of continuous star formation. To find the number of ionizing photons produced we need to know the mass of stars formed. To address this we present the number of ionizing photons formed per solar mass of the population ($\Ni/\mathrm{M_{tot}}$ weighted by $\xi(M)$) in \Cref{fig:Nphot_mass_perMsun}, for a population based on the SB13 IMF. By doing so, to know the number of photons produced by a Pop III population by time $t$ since initial star formation, you would only need to multiply by the mass of stars formed by that time. We are essentially representing continuous star formation by dividing the total photons produced by a population by the population's total mass, which, under the assumption of continuous and constant star formation, allows one to determine the number of ionizing photons produced at any time $t$ by the mass of stars formed at that time. This is a basic estimate and we encourage further studies to investigate the evolution of the ionizing photon production of Pop~III stars. All three ionizing photon species are included in \Cref{fig:Nphot_mass_perMsun}, for populations with and without rotation, and with higher overshooting. These values are also quoted in \Cref{mass_table_SB13}. This data is valuable for understanding the contribution of Pop III stars of different initial masses, weighted by the IMF, to reionization. We also present the total number of photons produced per solar mass by populations with different IMFs in \Cref{fig:Nphot_IMF_perMsun}. This is the same trend as is seen in the left panel of \Cref{fig:Nphot_effectsummary}, but now can be scaled based on the stellar mass of the population. These values are presented in \Cref{slope_table}. We note that the escape fraction is expected to vary with population mass  \citep{Kitayama2004,Wise2014}, so this needs be considered in scaling these values and determining the contribution to reionization of a given Pop~III population, the escape fraction is discussed further in \Cref{subsec:escape}.

\begin{table*}
 \centering   
 \caption{Total number of ionizing photons produced by individual stars over their lifetime ($\Ni$), for different initial masses. Headings indicate ionizing photon species H, He~I and He~II, and initial conditions where the top section gives non-rotating vs. rotating models, and the bottom section gives lower vs. higher overshooting models. These values are graphically presented in the left panels of \Cref{fig:Nphot_perstar,fig:Nphot_convection}.}
\begin{tabular}{|c|c|c|c|c|c|c|}
       \hline
    \noalign{\smallskip}
    $\mini$ (M$_\odot$)  & N$_\mathrm{H}$ (no rot) & N$_\mathrm{H}$ (rot) & N$_{\mathrm{HeI}}$ (no rot) & N$_{\mathrm{HeI}}$ (rot) & N$_{\mathrm{HeII}}$ (no rot) & N$_{\mathrm{HeII}}$ (rot)\\
    \noalign{\smallskip}
    \hline
    \noalign{\smallskip} 
      9  &   2.9697e+62  &   3.6921e+62  &   3.9789e+61  &   4.7302e+61  &   6.0183e+58  &   6.3006e+58\\
     12  &   6.3686e+62  &   7.3393e+62  &   1.1477e+62  &   1.2298e+62  &   4.2085e+59  &   3.9861e+59\\
     15  &   9.6272e+62  &   1.1304e+63  &   2.1015e+62  &   2.2958e+62  &   1.3917e+60  &   1.3631e+60\\
     20  &   1.5595e+63  &   1.8620e+63  &   4.1023e+62  &   4.4970e+62  &   4.9688e+60  &   4.7739e+60\\
     30  &   2.7315e+63  &   3.2348e+63  &   8.5739e+62  &   9.4340e+62  &   1.8505e+61  &   1.7676e+61\\
     40  &   4.0139e+63  &   4.4694e+63  &   1.3697e+63  &   1.4293e+63  &   3.9957e+61  &   3.6487e+61\\
     60  &   6.6986e+63  &   7.3097e+63  &   2.4910e+63  &   2.4663e+63  &   1.0103e+62  &   8.5209e+61\\
     85  &   9.9302e+63  &   1.0436e+64  &   3.9221e+63  &   3.7264e+63  &   1.9661e+62  &   1.5662e+62\\
    120  &   1.4816e+64  &   1.5774e+64  &   6.1039e+63  &   5.8296e+63  &   3.5737e+62  &   2.6896e+62\\
    \noalign{\smallskip}
    \hline
    \noalign{\smallskip}
            $\mini$ (M$_\odot$)  & N$_\mathrm{H}$ ($\alpha_{ov}\!=\!0.1$) & N$_\mathrm{H}$ ($\alpha_{ov}\!=\!0.3$) & N$_{\mathrm{HeI}}$ ($\alpha_{ov}\!=\!0.1$) & N$_{\mathrm{HeI}}$ ($\alpha_{ov}\!=\!0.3$) & N$_{\mathrm{HeII}}$ ($\alpha_{ov}\!=\!0.1$) & N$_{\mathrm{HeII}}$ ($\alpha_{ov}\!=\!0.3$)\\
    \noalign{\smallskip}
    \hline
    \noalign{\smallskip} 
      9  &   2.9697e+62  &   3.5357e+62  &   3.9789e+61  &   4.6342e+61  &   6.0183e+58  &   7.0159e+58\\
     15  &   9.6272e+62  &   1.4013e+63  &   2.1015e+62  &   2.9675e+62  &   1.3917e+60  &   1.8393e+60\\
     20  &   1.5595e+63  &   2.0425e+63  &   4.1023e+62  &   5.4138e+62  &   4.9688e+60  &   6.6130e+60\\
     30  &   2.7315e+63  &   3.3212e+63  &   8.5739e+62  &   1.0628e+63  &   1.8505e+61  &   2.4239e+61\\
     40  &   4.0139e+63  &   4.6039e+63  &   1.3697e+63  &   1.6100e+63  &   3.9957e+61  &   4.9812e+61\\
     60  &   6.6986e+63  &   7.2436e+63  &   2.4910e+63  &   2.7878e+63  &   1.0103e+62  &   1.2056e+62\\
     85  &   9.9302e+63  &   1.0629e+64  &   3.9221e+63  &   4.3961e+63  &   1.9661e+62  &   2.3507e+62\\
    120  &   1.4816e+64  &   1.6202e+64  &   6.1039e+63  &   7.0818e+63  &   3.5737e+62  &   4.5272e+62\\
    \noalign{\smallskip}
    \hline
    \noalign{\smallskip}
    \end{tabular}
    \label{initial_mass_table}

 \end{table*}

\begin{table*}
 \centering   
 \caption{Total number of ionizing photons produced per solar mass of the population, by stars of different initial masses, based on a population of IMF slope $\alpha\!=\!0.17$ ($\Ni\xi(M)/\mathrm{M_{tot}}$). Headings indicate ionizing photon species H, He~I and He~II, and initial conditions where the top section gives non-rotating vs. rotating models, and the bottom section gives lower vs. higher overshooting models. These values are graphically presented in \Cref{fig:Nphot_mass_perMsun}.}
\begin{tabular}{|c|c|c|c|c|c|c|}
    \hline
    \noalign{\smallskip}
    $\mini$ (M$_\odot$)  & N$_\mathrm{H}$ $\msun^{-1}$ (no rot) & N$_\mathrm{H}$ $\msun^{-1}$ (rot) & N$_{\mathrm{HeI}}$ $\msun^{-1}$ (no rot) & N$_{\mathrm{HeI}}$ $\msun^{-1}$ (rot) & N$_{\mathrm{HeII}}$ $\msun^{-1}$ (no rot) & N$_{\mathrm{HeII}}$ $\msun^{-1}$ (rot)\\
    \noalign{\smallskip}
    \hline
    \noalign{\smallskip} 
      9  &   5.9137e+58  &   7.3522e+58  &   7.9233e+57  &   9.4194e+57  &   1.1984e+55  &   1.2546e+55\\
     12  &   1.2077e+59  &   1.3917e+59  &   2.1763e+58  &   2.3320e+58  &   7.9805e+55  &   7.5586e+55\\
     15  &   1.7576e+59  &   2.0637e+59  &   3.8366e+58  &   4.1913e+58  &   2.5408e+56  &   2.4886e+56\\
     20  &   2.7112e+59  &   3.2371e+59  &   7.1321e+58  &   7.8183e+58  &   8.6384e+56  &   8.2996e+56\\
     30  &   4.4325e+59  &   5.2492e+59  &   1.3913e+59  &   1.5309e+59  &   3.0029e+57  &   2.8684e+57\\
     40  &   6.2027e+59  &   6.9066e+59  &   2.1166e+59  &   2.2086e+59  &   6.1746e+57  &   5.6384e+57\\
     60  &   9.6618e+59  &   1.0543e+60  &   3.5929e+59  &   3.5573e+59  &   1.4572e+58  &   1.2290e+58\\
     85  &   1.3499e+60  &   1.4187e+60  &   5.3319e+59  &   5.0658e+59  &   2.6728e+58  &   2.1292e+58\\
    120  &   1.8995e+60  &   2.0223e+60  &   7.8254e+59  &   7.4738e+59  &   4.5816e+58  &   3.4482e+58\\
    \noalign{\smallskip}
    \hline
    \noalign{\smallskip}
            $\mini$ (M$_\odot$)  & N$_\mathrm{H}$ $\msun^{-1}$ ($\alpha_{ov}\!=\!0.1$) & N$_\mathrm{H}$ $\msun^{-1}$ ($\alpha_{ov}\!=\!0.3$) & N$_{\mathrm{HeI}}$ $\msun^{-1}$ ($\alpha_{ov}\!=\!0.1$) & N$_{\mathrm{HeI}}$ $\msun^{-1}$ ($\alpha_{ov}\!=\!0.3$) & N$_{\mathrm{HeII}}$ $\msun^{-1}$ ($\alpha_{ov}\!=\!0.1$) & N$_{\mathrm{HeII}}$ $\msun^{-1}$ ($\alpha_{ov}\!=\!0.3$)\\
    \noalign{\smallskip}
    \hline    
    \noalign{\smallskip}
      9  &   5.9137e+58  &   7.0406e+58  &   7.9233e+57  &   9.2282e+57  &
   1.1984e+55  &   1.3971e+55\\
     15  &   1.7576e+59  &   2.6572e+59  &   3.8366e+58  &   5.6272e+58  &
   2.5408e+56  &   3.4878e+56\\
     20  &   2.7112e+59  &   3.7289e+59  &   7.1321e+58  &   9.8840e+58  &
   8.6384e+56  &   1.2073e+57\\
     30  &   4.4325e+59  &   5.7741e+59  &   1.3913e+59  &   1.8477e+59  &
   3.0029e+57  &   4.2141e+57\\
     40  &   6.2027e+59  &   7.4710e+59  &   2.1166e+59  &   2.6126e+59  &
   6.1746e+57  &   8.0832e+57\\
     60  &   9.6618e+59  &   1.1193e+60  &   3.5929e+59  &   4.3080e+59  &
   1.4572e+58  &   1.8630e+58\\
     85  &   1.3499e+60  &   1.5331e+60  &   5.3319e+59  &   6.3408e+59  &
   2.6728e+58  &   3.3906e+58\\
    120  &   1.8995e+60  &   2.2025e+60  &   7.8254e+59  &   9.6272e+59  &
   4.5816e+58  &   6.1545e+58\\
    \noalign{\smallskip}
    \hline    
    \noalign{\smallskip}    
    \end{tabular}
    \label{mass_table_SB13}

 \end{table*}

\begin{table*}
 \centering   
 \caption{Total number of ionizing photons produced per solar mass of the population, by populations of various IMF slopes ($\Npop/\mathrm{M_{tot}}$). Headings indicate ionizing photon species H, He~I and He~II, and initial conditions where the top section gives non-rotating vs. rotating models, and the bottom section gives lower vs. higher overshooting models. These values are graphically presented in \Cref{fig:Nphot_IMF_perMsun}. }
\begin{tabular}{|c|c|c|c|c|c|c|}
    \hline
    \noalign{\smallskip}
    IMF slope ($\alpha$)  & N$_\mathrm{H}$ $\msun^{-1}$ (no rot) & N$_\mathrm{H}$ $\msun^{-1}$ (rot) & N$_{\mathrm{HeI}}$ $\msun^{-1}$ (no rot) & N$_{\mathrm{HeI}}$ $\msun^{-1}$ (rot) & N$_{\mathrm{HeII}}$ $\msun^{-1}$ (no rot) & N$_{\mathrm{HeII}}$ $\msun^{-1}$ (rot)\\
    \noalign{\smallskip}
    \hline    
    \noalign{\smallskip}
  -1.00  &   1.1718e+62  &   1.2502e+62  &   4.6294e+61  &   4.4556e+61  &   2.3815e+60  &   1.8638e+60\\
  -0.50  &   1.1575e+62  &   1.2386e+62  &   4.5298e+61  &   4.3759e+61  &   2.2745e+60  &   1.7900e+60\\
   0.00  &   1.1360e+62  &   1.2211e+62  &   4.3854e+61  &   4.2588e+61  &   2.1306e+60  &   1.6886e+60\\
   0.17  &   1.1264e+62  &   1.2131e+62  &   4.3221e+61  &   4.2068e+61  &   2.0709e+60  &   1.6458e+60\\
   0.50  &   1.1033e+62  &   1.1937e+62  &   4.1727e+61  &   4.0826e+61  &   1.9363e+60  &   1.5482e+60\\
   1.00  &   1.0541e+62  &   1.1510e+62  &   3.8649e+61  &   3.8206e+61  &   1.6805e+60  &   1.3588e+60\\
   1.50  &   9.8479e+61  &   1.0887e+62  &   3.4484e+61  &   3.4555e+61  &   1.3669e+60  &   1.1207e+60\\
   2.00  &   8.9836e+61  &   1.0078e+62  &   2.9491e+61  &   3.0045e+61  &   1.0273e+60  &   8.5692e+59\\
   2.35  &   8.3354e+61  &   9.4503e+61  &   2.5874e+61  &   2.6696e+61  &   8.0269e+59  &   6.7897e+59\\
    \noalign{\smallskip}
    \hline
    \noalign{\smallskip}
        IMF slope ($\alpha$)  & N$_\mathrm{H}$ $\msun^{-1}$ ($\alpha_{ov}\!=\!0.1$) & N$_\mathrm{H}$ $\msun^{-1}$ ($\alpha_{ov}\!=\!0.3$) & N$_{\mathrm{HeI}}$ $\msun^{-1}$ ($\alpha_{ov}\!=\!0.1$) & N$_{\mathrm{HeI}}$ $\msun^{-1}$ ($\alpha_{ov}\!=\!0.3$) & N$_{\mathrm{HeII}}$ $\msun^{-1}$ ($\alpha_{ov}\!=\!0.1$) & N$_{\mathrm{HeII}}$ $\msun^{-1}$ ($\alpha_{ov}\!=\!0.3$)\\
    \noalign{\smallskip}
    \hline    
    \noalign{\smallskip}
  -1.00  &   1.1718e+62  &   1.2745e+62  &   4.6294e+61  &   5.2839e+61  &
   2.3815e+60  &   2.9342e+60\\
  -0.50  &   1.1575e+62  &   1.2625e+62  &   4.5298e+61  &   5.1727e+61  &
   2.2745e+60  &   2.7978e+60\\
   0.00  &   1.1360e+62  &   1.2456e+62  &   4.3854e+61  &   5.0164e+61  &
   2.1306e+60  &   2.6171e+60\\
   0.17  &   1.1264e+62  &   1.2381e+62  &   4.3221e+61  &   4.9490e+61  &
   2.0709e+60  &   2.5427e+60\\
   0.50  &   1.1033e+62  &   1.2205e+62  &   4.1727e+61  &   4.7916e+61  &
   1.9363e+60  &   2.3762e+60\\
   1.00  &   1.0541e+62  &   1.1831e+62  &   3.8649e+61  &   4.4709e+61  &
   1.6805e+60  &   2.0623e+60\\
   1.50  &   9.8479e+61  &   1.1297e+62  &   3.4484e+61  &   4.0389e+61  &
   1.3669e+60  &   1.6801e+60\\
   2.00  &   8.9836e+61  &   1.0605e+62  &   2.9491e+61  &   3.5192e+61  &
   1.0273e+60  &   1.2682e+60\\
   2.35  &   8.3354e+61  &   1.0064e+62  &   2.5874e+61  &   3.1397e+61  &
   8.0269e+59  &   9.9623e+59\\
    \noalign{\smallskip}
    \hline    
    \noalign{\smallskip}
    \end{tabular}
    \label{slope_table}

 \end{table*}   

\subsection{Initial Mass Function: Impact of minimum mass} \label{subsec:results:Mmin}

\begin{figure*}
    \centering
    \includegraphics[width=0.85\linewidth]{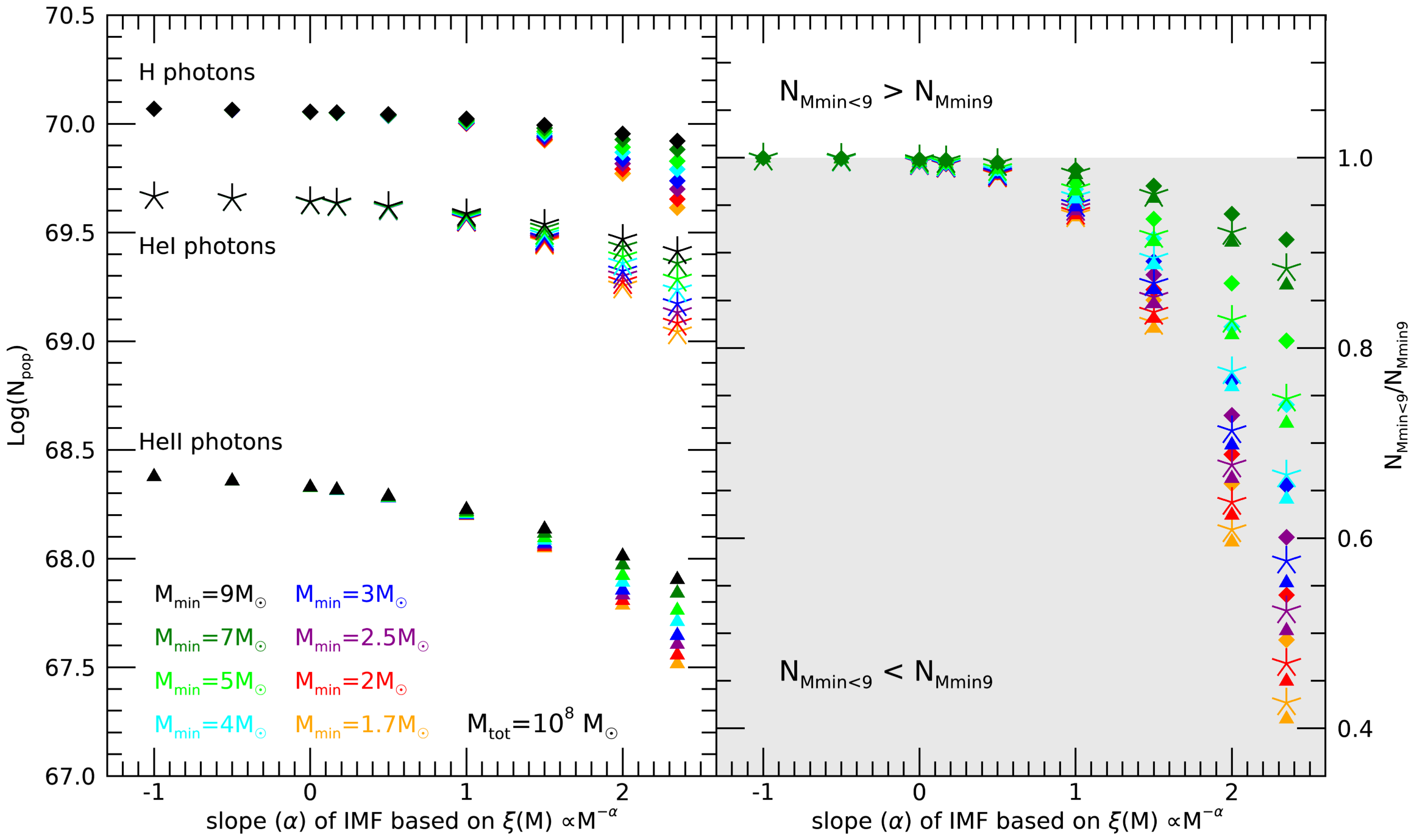}
    \caption{Impact of varying the minimum mass of the population, $\mmin$, on the ionizing photons produced by populations of different IMF slopes, $\alpha$, based on the relation $\xi(M) \propto M^{-\alpha}$. All stars are non-rotating with convective overshooting of $\alpha_{ov}=0.1$, and the total mass of the population is $\mathrm{M_{tot}}=10^8\,\msun$ as indicated in the figure. Colours indicate the minimum mass of the population, and H, He~I and He~II ionizing photons are indicated by different symbols. \textit{Left:} Total ionizing photons produced. \textit{Right:} Ratio of ionizing photons produced by populations of lower minimum mass to our fiducial population with $\mmin=9\,\msun$, for H, He~I and He~II ionizing photons.}
    \label{fig:NpopvsMmin}
\end{figure*}

Up to now we have assumed a minimum mass of the population of $\mmin=9\,\msun$. We will now investigate how the ionizing photon production of the population is impacted by decreasing $\mmin$. To compare how changing $\mmin$ impacts the total ionizing photons produced by the population, we reevaluate the number of stars at each initial mass in a population of $\mathrm{M_{tot}}=10^8\,\msun$. The results of this investigation are presented in \Cref{fig:NpopvsMmin}. It is clear from the plot that the impact of varying $\mmin$ is highly dependent on the IMF slope. For IMF slopes of $\alpha<1$ the ionizing photon production of the population remains largely unchanged as the minimum mass decreases. However, for IMF slopes $\geq\,$1 we begin to see a significant decrease to the total ionizing photons produced. For example, we find a reduction of $\sim\,$50\% to H ionizing photons produced by a population with Salpeter IMF when $\mmin$ is decreased from 9$\,\msun$ to 1.7$\,\msun$. This result is unsurprising given that steeper IMFs significantly increase the fraction of low-mass stars for a population of fixed total mass. Decreasing the minimum mass in these steeper IMF populations thus reduces the number of ionizing photons produced because lower-mass models are cooler and less luminous (see Figure 1, \citealt{Murphy2021}), therefore, they have much lower ionizing photon production rates.

Since the Pop~III IMF is expected to be top-heavy \protect\citep{Greif2011,StacyBromm2013,Susa2014,Hirano2014,Hirano2015,Stacy2016,Jerabkova2018,Wollenberg2020}, a variation of minimum mass in the range $1.7-9\,\msun$ may have little effect on the ionizing photon production of the population. However, it is important to keep this effect in mind as research on the primordial IMF continues. We also note that changing the minimum mass of the population may have a significant impact on the timescale of reionization. As we discussed in \Cref{subsec:results:IMFevol} more top-heavy IMF populations produce their ionizing photons faster than populations with a Salpeter IMF slope, since more massive models have shorter lifetimes. Reducing the minimum mass of the population would delay the production of ionizing photons by a Salpeter IMF population even further, since the lifetimes of these low-mass stars can be significantly longer. A $1.7\,\msun$ Pop~III star has a lifetime of $\sim$ a billion years, compared to the 20$\,$Myr lifetime of a 9$\,\msun$ star. Furthermore, the escape fraction is expected to change rapidly as a function of redshift (e.g. \citealt{HaardtMadau2012}), so not only will decreasing the minimum mass increase the timescale for ionizing photon production, but the escape fraction may be very different, which will impact the number of ionizing photons that reach the IGM.

\subsection{Initial Mass Function: Impact of maximum mass} \label{subsec:results:Mmax}

\begin{figure*}
    \centering
    \includegraphics[width=0.85\linewidth]{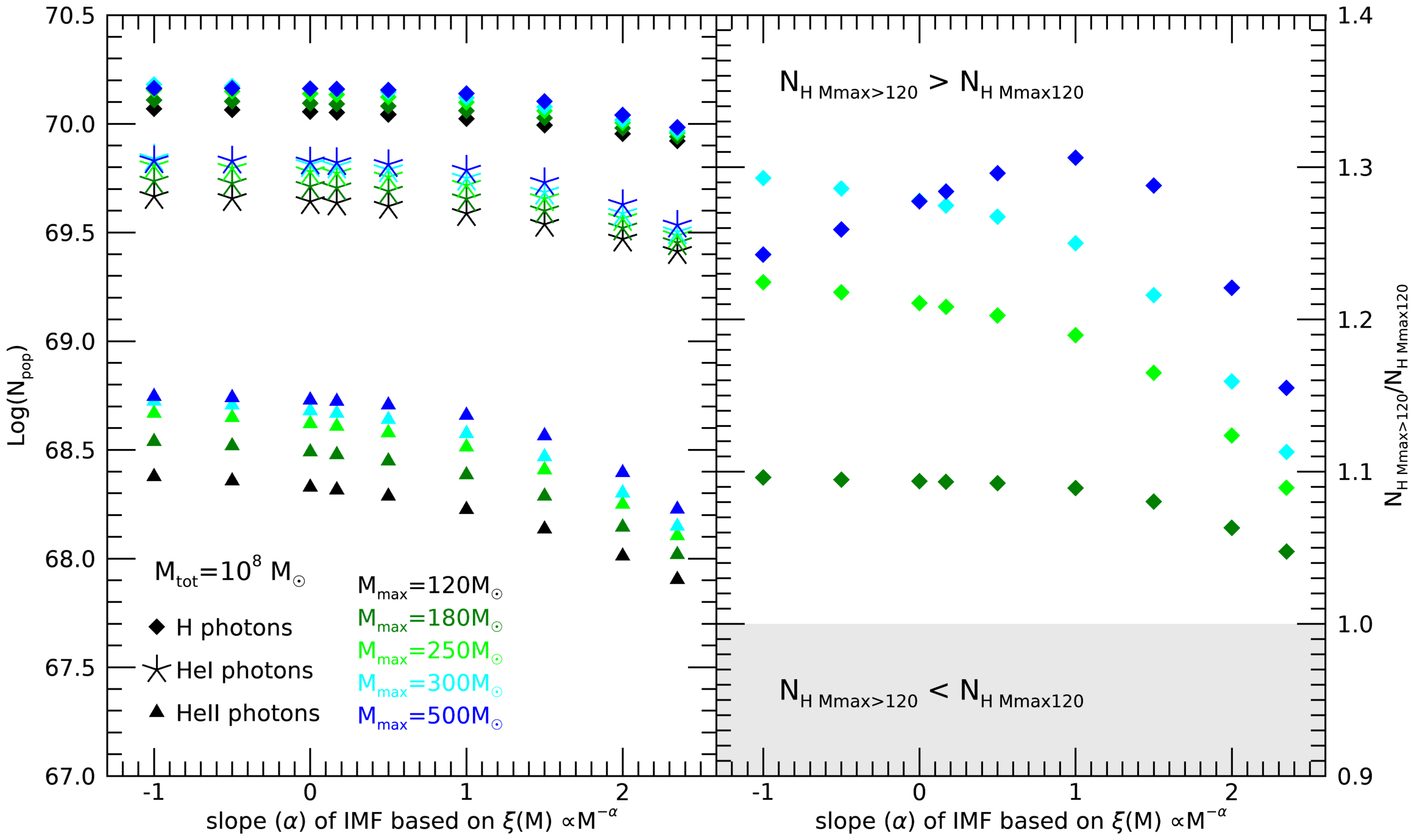}
    \caption{Impact of varying the maximum mass of the population, $\mmax$, on the ionizing photons produced by populations of different IMF slopes, $\alpha$, based on the relation $\xi(M) \propto M^{-\alpha}$. All stars are non-rotating with convective overshooting of $\alpha_{ov}=0.1$, and the total mass of the population is $\mathrm{M_{tot}}=10^8\,\msun$ as indicated in the figure. Colours indicate the maximum mass of the population, and H, He~I and He~II ionizing photons are indicated by different symbols. \textit{Left:} Total ionizing photons produced. \textit{Right:} Ratio of H ionizing photons produced by populations of higher maximum mass to our fiducial population with $\mmax=120\,\msun$.}
    \label{fig:NpopvsMmax}
\end{figure*}

We now turn our attention to the upper mass limit of the population, and investigate how the ionizing photon production of the population is impacted by increasing the maximum mass, $\mmax$, using our original minimum mass value of $\mmin=9\,\msun$. We use the recently computed zero-metallicity very massive star models from Martinet et al., in prep, with initial masses $\mini$\,=\,180,\,250,\,300,\,500$\,\msun$. Similarly to \Cref{subsec:results:Mmin}, we vary $\mmax$ and reevaluate the number of stars in the population, keeping the total stellar mass constant with $\mathrm{M_{tot}}=10^8\,\msun$. \Cref{fig:NpopvsMmax} shows how varying the maximum mass affects the total ionizing photons produced by a zero-metallicity population. It can be clearly seen from \Cref{fig:NpopvsMmax} that increasing the maximum mass from 120$\,\msun$ increases the total ionizing photons produced by the population for all IMF slopes shown here. We know from \Cref{subsec:results:rotationNi} that total ionizing photons produced, $\Ni$, increases with increasing initial mass, and the same is true for masses $>\!120\,\msun$. However, from \Cref{fig:NpopvsMmax} we see an interesting change in trend for the $\mmax=500\,\msun$ case. 

\begin{figure}
    \centering
    \includegraphics[width=\linewidth]{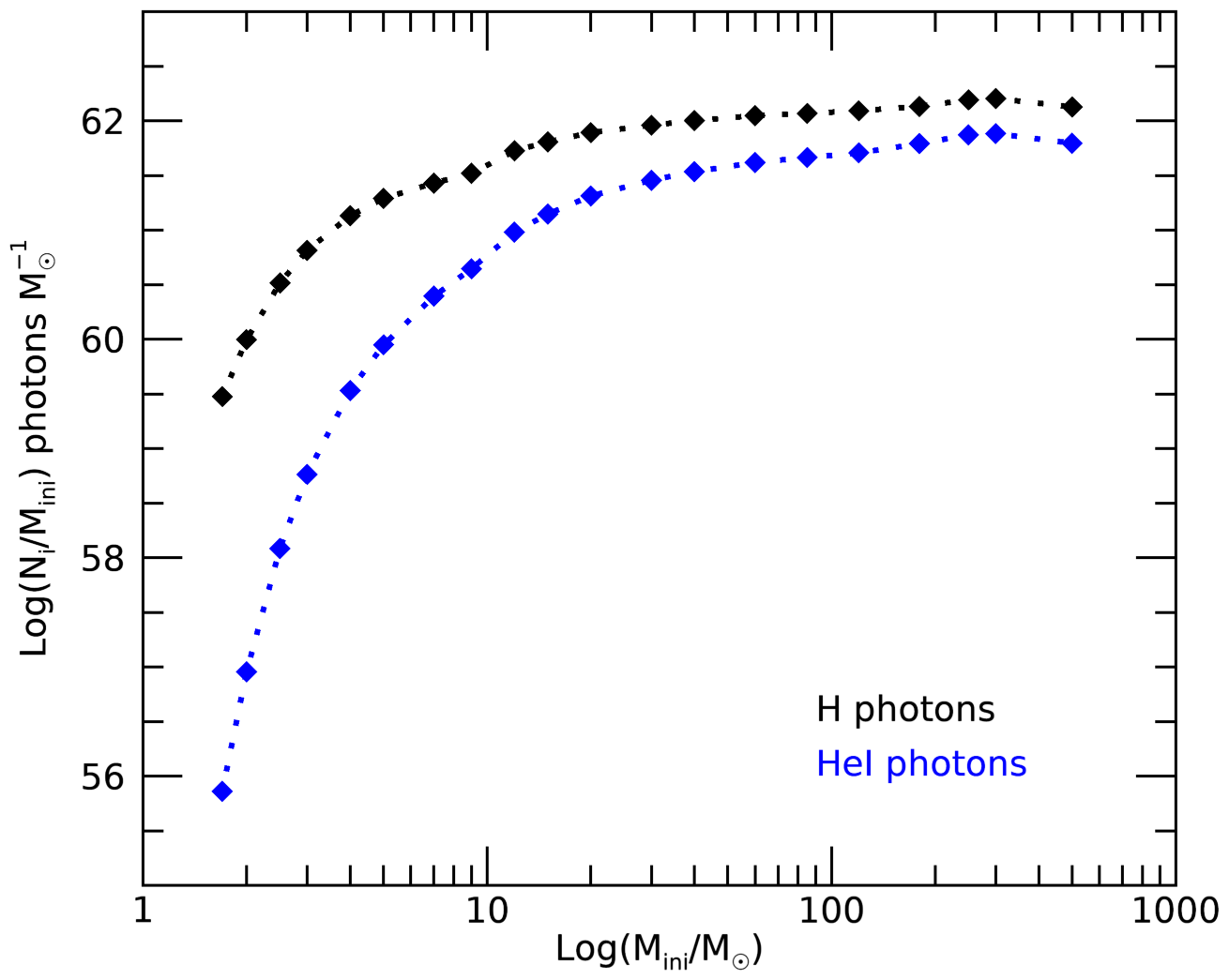}
    \caption{Total ionizing photons ($\Ni$) produced by models in the mass range $1.7\,\msun\!\leq\!\mini\!\leq\!500\,\msun$, divided by their initial mass ($\mini$), for H ionizing photons in black, and He~I ionizing photons in blue.}
    \label{fig:NiperMini}
\end{figure}

Looking at the right panel it can be observed that when increasing $\mmax$ to $500\,\msun$, the maximum increase in emission of H ionizing photons occurs at an IMF slope of $\alpha=1$. For $\alpha\leq0$ the increase to H ionizing photon produced is in fact higher for $\mmax=300\,\msun$ than $\mmax=500\,\msun$. We have found that this occurs because, despite the 500$\,\msun$ model producing more ionizing photons than the 300$\,\msun$ model, it produces less ionizing photons per solar mass than the 300$\,\msun$. This can be observed from \Cref{fig:NiperMini}, where we plot the total ionizing photons produced by individual stars divided by their initial mass ($\Ni/\mini$). From this figure we see that the total H ionizing photons produced per initial mass increases rapidly moving from intermediate mass models ($1.7\,\msun\!\leq\!\mini\!\leq\!9\,\msun$) to massive models $\mini\geq9\,\msun$, showing that massive models will play a dominant role in ionizing photon production. The values for $\log(\Ni/\mini)$ peak at 300$\,\msun$ and decrease moving to 500$\,\msun$. This means for instance that two 250$\,\msun$ stars would produce more ionizing photons than one 500$\,\msun$ star. Since the stellar mass of the population is kept constant, when 500$\,\msun$ models are included the total ionizing photons produced decrease compared to the population with $\mmax=300\,\msun$, or $\mmax=250\,\msun$.

This decrease in $\log(\Ni/\mini)$ for 500$\,\msun$ models occurs due to an inflation of their envelope due to their high Eddington factors (see also \protect\citealt{Sanyal2017} for Pop~III models). This inflation causes their temperature to decrease, thus impacting their ionizing photon production. It is clear that the ionizing photons produced per solar mass is crucial in understanding how stars of different initial mass will contribute to the ionizing photon production of the population. We will further discuss the impact of envelope inflation in \Cref{subsec:inflation}.

From \Cref{fig:NpopvsMmax} we find that varying the maximum mass of the population can increase the total H ionizing photons produced by up to $\sim\,$30\% for $\mmax=500\,\msun$ compared to $\mmax=120\,\msun$. We have found that for a SB13 IMF ($\alpha=0.17$) the total H ionizing photons produced increases by 28.4\% when $\mmax$ is increased from 120$\,\msun$ to 500$\,\msun$, while for a Salpeter IMF ($\alpha=2.35$) this increase is 15.5\%. Given that the minimum and maximum initial masses of Pop~III stars are still debated in cosmological simulations (see Table 1, \citealt{Stacy2016}), this significant shift in total ionizing photons produced as maximum mass varies should be considered in future studies of how the first stars contributed to reionization.

We note here that we have considered a constant IMF slope across the full mass range of the population. Hydrodynamical simulations of Pop~III star formation predict a more complex distribution however (e.g. Figure 17 of \citealt{Hirano2015}), with a non-monotonic IMF. While a top-heavy IMF is predicted, the characteristic mass is not yet certain and the nature of the IMF above this characteristic mass is yet to be constrained. Therefore, while the maximum mass of the population is expected to be $>\!120\,\msun$ the IMF slope may change for very massive stars, and thus we advise the reader to carefully consider the nature of the IMF slope in different mass ranges, as well as the minimum and maximum masses in predicting Pop~III ionizing photon production for a given IMF.



\section{Discussion} \label{sec:discussion}

\subsection{Impact of mixing processes: \textsc{\small SNAPSHOT} models} \label{subsec:snapshot}
\begin{figure}
    \centering
    \includegraphics[width=\linewidth]{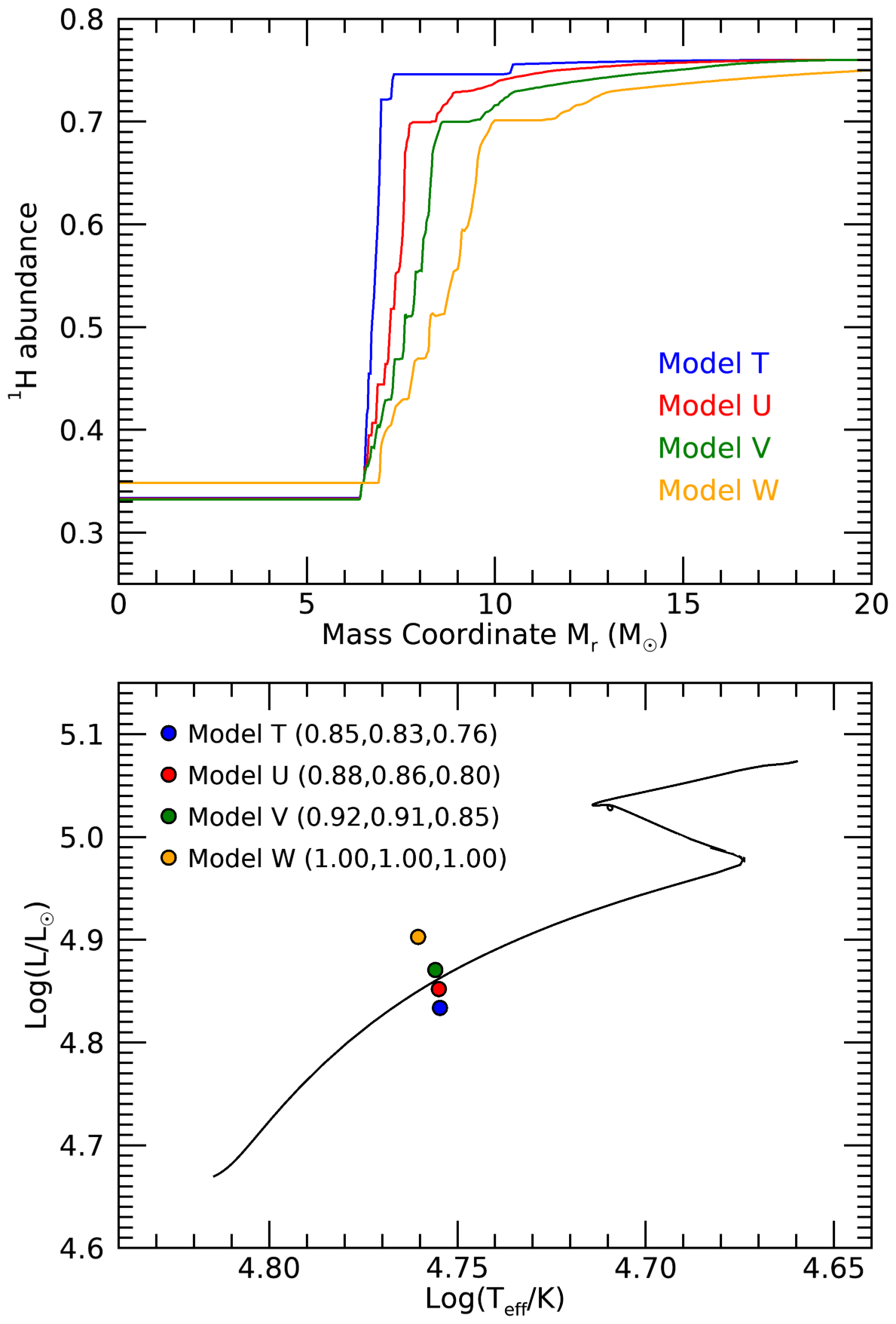}
    \caption{\textit{Upper:} Internal hydrogen abundance profiles of four 20~$\msun$ \textsc{\small SNAPSHOT} stellar models at the middle of the MS phase.  \textit{Lower:} Location in the HR diagram of the models from the upper panel. For reference, we also plot the evolutionary track of a non-rotating 20$\,\msun$ computed with MESA. For each \textsc{\small SNAPSHOT} model T, U, V, we quote the fraction of ionizing photons produced relative to model W in order of H, He~I, He~II respectively.}
    \label{fig:snapshot}
\end{figure}

The luminosity, effective temperature and ionizing photon production rate are ultimately set by the internal abundance profile. To test the effects of the uncertainties in the internal hydrogen abundance profile on the production of ionizing photons produced in our models, we compute several \textsc{\small SNAPSHOT} stellar structure models using a the method described in \citep{Farrell2020_snapshot} with the \textsc{MESA} stellar evolution code \citep{Paxton2011, Paxton2013, Paxton2015}. The \textsc{\small SNAPSHOT} models allow us to test the effect of arbitrary internal abundance profiles on the luminosity and effective temperature without relying on results based on specific prescriptions for mixing. While the abundance profiles are artificial and generated by hand, they mimic the effects produced by physical processes such as rotation, convection, semi-convection and other mixing processes that may have effects but are not accounted for in these models. We find that slight variations in the internal hydrogen abundance profile can affect the rate of ionizing photons by up to 20\% during the middle of the MS and by 30\% at the end of the MS. 
Fig. \ref{fig:snapshot} shows the internal hydrogen abundance profile for four 20$\,\msun$ \textsc{\small SNAPSHOT} stellar models at the middle of the MS, as well as their location in the HR diagram. Quoted in the figure are the fractions, for models T, U, V, of ionizing photons produced relative to model W, for H, He~I, and He~II respectively. This gives an idea of the extent to which moderate adjustments to the abundance profile may impact surface properties, and subsequently the number of ionizing photons produced. We conclude that moderate variations of the sort shown in \Cref{fig:snapshot} can have a moderate effect on the overall result for ionizing photon production.

\subsection{Impact of detailed stellar atmosphere modelling} \label{subsec:Sch02}
\begin{figure*}
    \centering
    \includegraphics[width=0.85\linewidth]{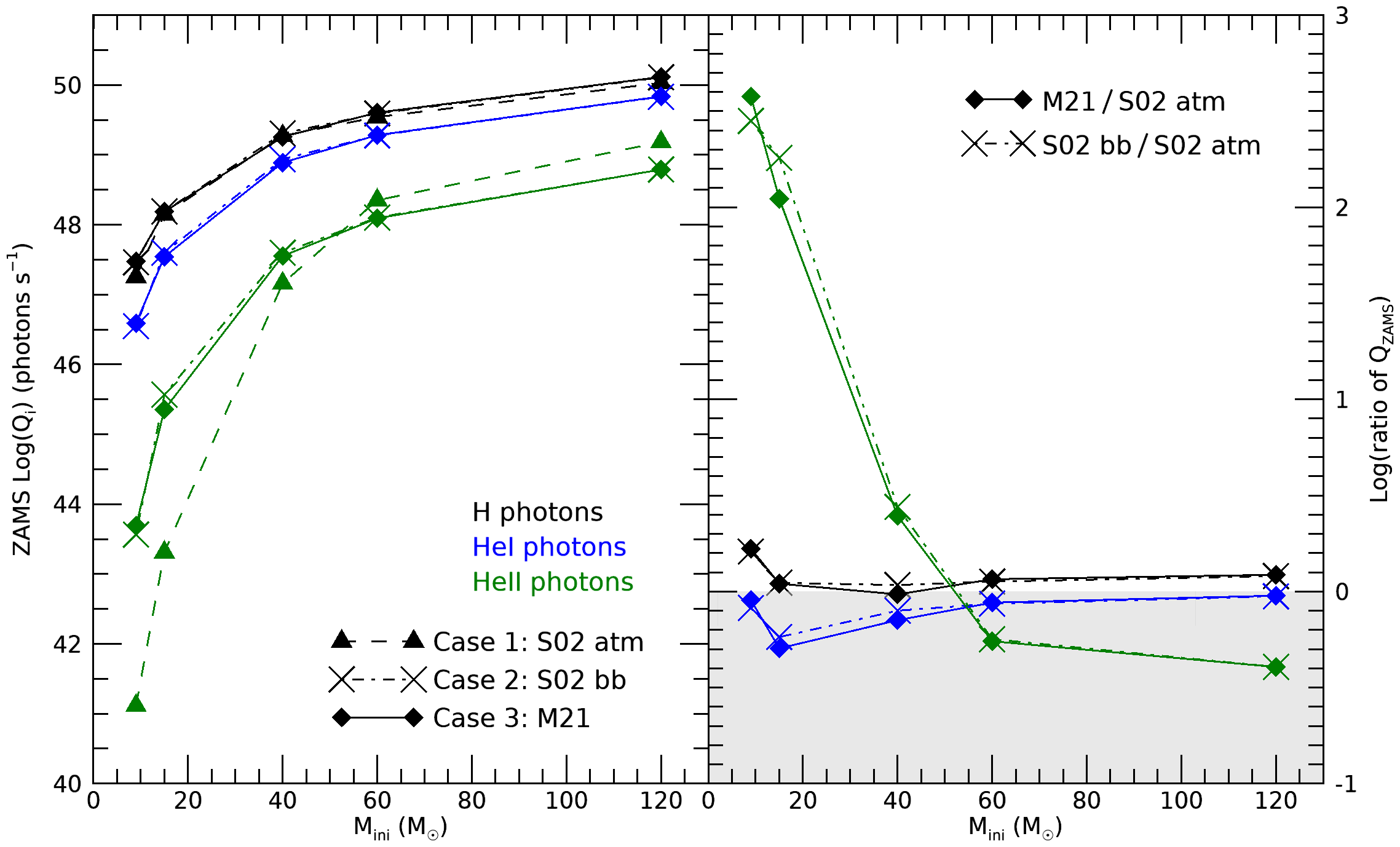}
    \caption{\textit{Left:} Comparison to \citet{Schaerer2002} (S02) ionizing photon production rates ($\Qi$) for non-rotating ZAMS models. Triangle symbols and dashed lines indicate the values from Table 3 in S02. Cross symbols and dashed dotted lines indicate values using $L$ and $\teff$ values from S02 but computed using our blackbody method described in \Cref{sec:methods}. Diamond symbols and solid lines indicate ionizing photon production rates from non-rotating models in this work on the ZAMS. \textit{Right:} Ratio of $\Qi$ from this work to S02 values (diamonds and solid lines), and ratio of $\Qi$ from S02 calculated using blackbody to S02 values (crosses and dashed dotted lines). Results are logarithmically scaled. Colours indicate ionizing photon species as shown in the left panel.}
    \label{fig:compareS02}
\end{figure*}

To determine how much our results are impacted by using a blackbody rather than modelling the atmosphere we can compare to \cite{Schaerer2002} (S02). Since the stellar evolution models used are not the same we choose three cases to compare our findings with the S02 values. Case 1 is the values of $\QH$, $\QHeI$, $\QHeII$ taken directly from S02 Table 3, which are calculated from a modelled atmosphere. These values are the ionizing photon production rates at the ZAMS for each stellar evolution model. We select the values for initial masses in the range $9\,\msun\!\leq\!\mini\!\leq\!120\,\msun$, of which four masses overlap with the initial masses included in this work, these are $\mini=9,15,60,120\,\msun$. Case 2 takes the $L$ and $\teff$ values from the same table in S02, but uses our method (see \Cref{sec:methods}) to calculate $\QH$, $\QHeI$ and $\QHeII$ through a blackbody fit, thus tests the impact of not modelling the atmosphere. Case 3 is then the ZAMS $\Qi$ values from this work for our non-rotating models. The comparison of these three cases is shown in the left panel of \Cref{fig:compareS02}, while ratios of case 2 to case 1, and case 3 to case 1 are shown in the right panel. For each case we linearly interpolate the values so that we can compare all three cases for the full mass range $9\,\msun\!\leq\!\mini\!\leq\!120\,\msun$. 

It is clear from the figure that the most significant impact of not modelling the atmosphere is the overestimation of He~II photons at low initial masses. In particular, if we look at the right panel of \Cref{fig:compareS02}, we see that He~II photons are overestimated for initial masses $\mini\!<\!40\,\msun$ when we use a blackbody fit. Referring back to the rightmost panel of \Cref{fig:NphotIMF}, we see that for all IMFs with slopes $\alpha\!\leq\!2.35$ He~II ionizing photon production is dominated by models of initial masses $\mini\!\geq\!40\,\msun$. In this mass range we have found that our blackbody fit reproduces the ionizing photon production rates with errors of less than 0.5\,dex (see right panel \Cref{fig:compareS02}). This is not necessarily the case for larger IMF slopes of $>\!2.35$ where we expect that He~II ionizing photon production is dominated by initial masses of $<\!40\,\msun$, however, simulations predict that there were more massive stars in the early universe \citep{Turk2009,Clark2011TheProtostars,Greif2012,StacyBromm2013,Hirano2014,Hirano2015} and it is generally accepted that the Pop III IMF will have a lower slope than the Salpeter IMF. 

The numbers of H and He~I ionizing photons are well reproduced with a blackbody approximation within $\lesssim0.2\,$dex. Therefore, our comparison to the ionizing photon values presented in \cite{Schaerer2002} has shown that our method of using a blackbody fit to determine ionizing photon production, rather than modelling the atmosphere, is a reasonable approximation. It is important to note here that this is for ZAMS values, therefore, we may see more absorption of photons at later times if surface metallicity increases, however this is not expected for our models. 

\subsection{Impact of internal magnetic fields} \label{subsec:Yoon}
It is also useful to compare to \cite{Yoon2012EvolutionFields} (Y12) to study how differences in stellar evolution modelling will impact ionizing photon production. Internal magnetic fields were included in the Y12 Pop~III stellar models, assuming a Taylor-Spruit dynamo \citep{Spruit2002}. This allowed for a different approach to the effects of rotation in these stars, where chemically homogeneous evolution (CHE) is more easily achieved (see discussion in \citealt{Groh2019Grids0.0004}). The evolution of stellar surface properties is strongly impacted by CHE, in particular CHE leads to much higher surface temperatures. Therefore, we expect that the inclusion of magnetic fields and subsequent CHE will increase ionizing photon production. To test this we have selected four models from Y12 rotating models, for the same initial mass, with similar initial rotational velocities to our models of $\vini\!=\!0.4\vcrit$. These include m15vk04 ($\vini\!=\!0.43\vcrit$), m20vk04 ($\vini\!=\!0.44\vcrit$), m30vk04 ($\vini\!=\!0.47\vcrit$), and m60vk03 ($\vini\!=\!0.39\vcrit$), of initial masses $\mini\!=\!15,20,30,60\,\msun$, and we direct the reader to their paper for further details on initial parameters. We compare the total ionizing photons produced by these models (see Y12 Table 3) to the total ionizing photons produced by our rotating and non-rotating models of the same initial masses. 

\begin{figure*}
    \centering
    \includegraphics[width=0.85\linewidth]{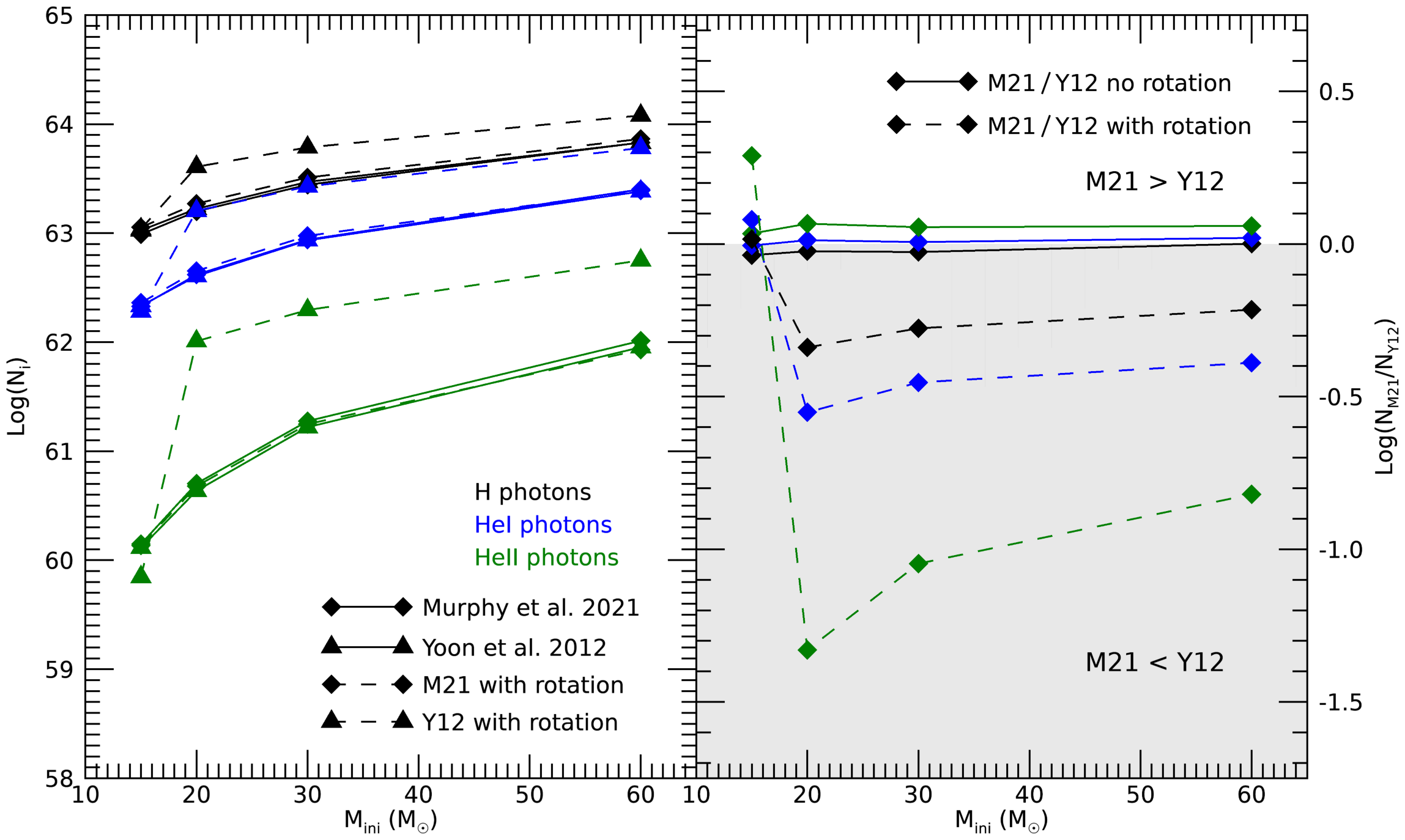}
    \caption{\textit{Left:} Comparison to ionizing photons produced over lifetime (N$_i$) by \citet{Yoon2012EvolutionFields} (Y12) models to values in this work. Rotating models are shown by dashed lines, and selected for having similar initial rotational velocities to our rotating models in \citet{Murphy2021} (M21), while solid lines show ionizing photons produced by non-rotating models. Triangle symbols indicate Y12 model values, and diamond symbols indicate ionizing photons produced by M21 models which are presented in this work. \textit{Right:} Ratio of N$_i$ from this work to Y12 values. Colours indicate ionizing photon species as shown in the left panel.}
    \label{fig:compareY12}
\end{figure*}

The results of this comparison are presented in \Cref{fig:compareY12}. As expected we see that the ionizing photon output is higher for rotating Y12 models, with the exception of the 15$\,\msun$ case where CHE is not achieved. From the right panel of \Cref{fig:compareY12}, we see that He~II photons are the species most impacted by CHE. This is because He~II ionizing photon production is dominated by surface temperature effects (see \Cref{subsec:results:rotationQi}). The average increase to He~II ionizing photons produced with CHE (for $\mini\!=\!20,30,60\,\msun$) is 1.02\,dex, while for He~I ionizing photons the average increase is 0.46\,dex, and for H ionizing photons the average increase is 0.27\,dex. These results tell us that for our massive rotators ($\mini \geq 20\,\msun$), we can expect to see an increase in ionizing photon production if magnetic fields are included. Looking at the right panel of \Cref{fig:compareY12}, we also see a small increase ($<\!0.1\,$dex) in He~II photons when we compare our non-rotating models to the Y12 non-rotators. This is due to higher surface temperatures in our models. As with this work, \cite{Yoon2012EvolutionFields} assumed blackbody radiation. An important consequence of CHE is the increase to surface metallicity, so the increase in ionizing photon production with the inclusion of magnetic fields may vary with radiative transfer modelling of the atmosphere. Therefore, further work with modelling of the atmosphere is required to accurately determine the impact of magnetic fields on the production of ionizing photons by Pop III stars.

\subsection{Impact of envelope inflation} \label{subsec:inflation}

Hot massive stars have envelopes in which the transport of energy occurs mainly through radiation. \protect\citet{Grafener2012} discussed the possibility of formation of inflated stellar envelopes when stars have high Eddington parameters, $\gtrsim\,$0.3 (see also \protect\citealt{Ishii1999,Petrovic2006} for further discussion on envelope inflation), which is likely for very massive stars due to their high luminosities. In these works the inflation of the envelope is mainly triggered by the presence of a sudden increase in opacity due to Fe (the so-called Fe bump). \protect\citet{Sanyal2017} presented numerical stellar evolution models for a range of metallicities, showing that the nature of the inflated envelopes is sensitive to the Fe content. They also showed that for Pop~III models, the presence of inflated envelopes during the MS should occur for $\mini\gtrsim150\,\msun$. If these stars do experience inflation of their envelopes, their surface temperature will subsequently decrease, which in turn will decrease their ionizing photon production rates.

Based on these previous studies, we do not expect Pop~III stars to experience substantial envelope inflation during the MS for $\mini \lesssim 150\,\msun$, which is consistent with our model grid \citep{Murphy2021}. Inflation may be important for the very massive star models, and based on the one-dimensional energy transport prescriptions available we estimate a decrease of around $\sim20\%$ for the number of hydrogen ionizing photons coming from individual stars above 150$\,\msun$. This is a rough estimate, obtained by comparing the number of photons produced per $\msun$ for the 300$\,\msun$ model to the 500$\,\msun$ model. As we discussed in \Cref{subsec:results:Mmax}, the impact will be lower than 20\% for a population of fixed mass and reasonable IMF, since the majority of the stars will have $\mini < 150\,\msun$.

It is worth noting that ultimately the structure and radial extension of radiatively-dominated envelopes is challenging to accurately model in 1-D. We encourage the development of 3-D radiation hydrodynamic simulations for Pop~III stellar envelopes, similarly to the study of solar-metallicity models from \protect\citet{Jiang2015}.

\subsection{Supernovae, binaries and supermassive Pop~III stars} \label{subsec:SNe}
We note that in this work we have studied the ionizing photon production during the lifetime of individual stars, and integrated this to determine the ionizing photon production of a stellar population. We have therefore omitted the impact of explosive events such as supernovae in the contribution to ionization of the first stellar populations. In addition, our models are based on single stars. Yet, we stress that the impact of the physical effects discussed here, such as rotation and convective overshooting, will also hold for binary stars. \citet{Berzin2021} discussed the contribution of stripped stars to reionization. These authors found that the contribution of stripped stars compared to single stars reduces moving towards lower metallicities, but stripped stars still have a noticeable effect in the spectral energy distribution of galaxies in the early Universe. The relative contribution of binary stars to the production of ionizing photons depend on their initial orbital separation. The statistics of Pop~III binaries were investigated in \cite{Liu2021}, which showed that relatively close binaries (semi major axis $< 100$\,au) tend to be rare in Pop~III systems. This is in contrast with many previous results \citep{Kinugawa2014,Hartwig2016,Belczynski2017,LiuBromm2020b} and tends to weaken the importance of close binary evolution for these Pop~III stars. If close binary evolution is rare for Pop~III stars, we expect our results for the ionizing photon production of Pop~III stellar populations to be largely unchanged by the inclusion of binaries. Given the caveats associated with Pop III star formation, we encourage further research to investigate Pop~III binaries and their impact on ionization. 

Recent studies have provided new insights into the nature of Pop~III supermassive stars (SMSs) and their role in the formation of the first quasars \citep{Hosokawa2013,Umeda2016,Woods2017,Haemmerle2018,Haemmerle2019,Haemmerle2021}, these objects are expected to contribute significantly to ionization, so should be included in further studies alongside the mass range of stars discussed in this work. The detectability of such objects with JWST has been investigated in \cite{Martins2020}, showing that SMSs should be detectable if they are luminous and relatively cool. Additionally, for a narrow range of initial masses it had been found that SMSs may explode as general relativity supernovae (GRSNe) \citep{Chen2014}. The detectability of such enormous explosive events by JWST and Galaxy and Reionization EXplorer (G-REX) was studied recently in \cite{Moriya2021}. They found that GRSNe would have long plateau phases that appear as persistent sources at high redshifts, but would be distinguishable from high redshift galaxies. If SMSs and GRSNe are found with these new facilities in the coming years, then it is likely that they will significantly affect reionization at high redshifts.

\subsection{Escape rates of ionizing photons} \label{subsec:escape}

The contribution of the first stars to cosmic reionization not only depends on ionizing photons produced, but also on the escape fraction (\fesc) of these photons from the minihalo or host galaxy. Simulations of the early Universe have been used to study the contribution of the first stellar populations to hydrogen reionization \citep{GnedinOstriker1997,Sokasian2004,Whalen2004,Kitayama2004,McQuinn2007,HaardtMadau2012,Wise2014,Finlator2018,Katz2019} and can reproduce hydrogen reionization up to $z\!=\!6$. However, to do so they require escape fractions of at least 20\%. Some studies have shown that the first galaxies had high escape fractions with \cite{Whalen2004} suggesting $\fesc>0.95$, and \cite{WiseCen2009} finding escape fractions of $0.25\!<\!\fesc\!<\!0.8$ for galaxies of masses greater than $10^7\,\msun$ with a top-heavy initial mass function (IMF). \cite{Alvarez2006} also showed high escape fractions of up to 90\% when stars of masses $80\msun\!<\!\mini\!<\!500\msun$ are included. More recent work \citep{Wise2014} has shown that the escape fraction is inversely dependent on halo mass with $\fesc \!\sim\! 0.5$ for halo masses $<2\times10^7\,\msun$ and escape fractions as low as $\fesc \!\sim\! 0.05$ for halo masses $>2\times10^8\,\msun$, which is similar to results found in \cite{Kitayama2004} showing that halos of mass $<10^6\,\msun$ have escape fractions of more than 80\% while the escape fraction for higher mass halos ($>\!10^7\,\msun$) is essentially zero. This inverse relationship of escape fraction and halo mass was further supported by the Renaissance simulations \citep{Xu2016}. As work to constrain the escape fraction of ionizing photons continues, it is important to keep this limit in mind in determining how the first stars contributed to reionization. Through simultaneous efforts to predict the ionizing photon contribution of Pop~III stars and the escape rates of photons from these populations we move closer to understanding the epoch of reionization, and prepare for prospective detections from facilities such as JWST.


\section{Conclusions} \label{sec:conclusions}

Here we have investigated the ionizing photon production rates and total ionizing photons produced by our Geneva Pop~III model grid. We have presented analytical fits for the ionizing photons produced by non-rotating zero-metallicity models in the mass range $1.7\,\msun\!\leq\!\mini\!\leq\!500\,\msun$. We have discussed the results for individual models and explained the impact of initial mass, rotation, and convective overshooting. We have analysed the ionizing photon production of populations of these stars for a number of different IMFs, and showed how this evolves over time. We have also compared our results to previous works on ionizing photon production of Pop~III stellar evolution models, \cite{Schaerer2002} and \cite{Yoon2012EvolutionFields}. The following points summarise our findings.

\begin{itemize}
    \item The total number of ionizing photons produced over the lifetime of individual stars increases with increasing initial mass despite shorter lifetimes at higher initial masses. This is due to the higher luminosity and surface temperatures of more massive models which drive higher ionizing photon production rates.
    \item Rotation impacts the total ionizing photons produced by up to 25\% for the initial rotational velocity considered here. The most significant impact is higher H ionizing photon production for less massive models and lower He~II ionizing photon production for more massive models. The difference in rotational effects for different ionizing photon species reflects the dominant surface property in each case. H ionizing photons are dominated by changes in luminosity, while He~II ionizing photons are dominated by changes in surface temperature.
    \item Higher convective overshooting increases the total number of ionizing photons produced for all species at all initial masses considered here. Increases in luminosity and surface temperature contribute to higher ionizing photon production rates but increased lifetimes is the dominant factor in increasing total ionizing photons produced with higher convective overshooting.
    \item Ionizing photon production increases with decreasing IMF slope, because more top-heavy IMFs are dominated by higher initial masses which produce more ionizing photons. This variation depends on the IMF slopes considered, with a decrease of 26\% comparing the non-rotating $\alpha_{ov}=0.1$ SB13 IMF population to the Salpeter IMF population.
    \item Along with producing less ionizing photons than more top-heavy IMF populations, populations of higher IMF slope, $\alpha$, take longer to produce ionizing photons since they are dominated by less massive models with longer lifetimes.
    \item Varying the minimum mass, $\mmin$, of the population decreases the total H ionizing photons produced for populations with IMF slopes of $\alpha\geq1$, by up to $\sim\,$50\% for $\mmin=1.7\,\msun$ compared to $\mmin=9\,\msun$ where $\alpha=2.35$.
    \item Varying the maximum mass, $\mmax$, of the population increases the total H ionizing photons produced for all IMF slopes considered here, by up to $\sim\,$30\% for $\mmax=500\,\msun$ compared to $\mmax=120\,\msun$ where $\alpha=1$.
    \item Our modifications to the H abundance profile using the \textsc{\small SNAPSHOT} method impact the ionizing photon production rates of a 20$\,\msun$ model by up to 20\% halfway through the MS and 30\% at the end of the MS.
    \item Through comparing our results to \cite{Schaerer2002}, we have found that our approach of using a blackbody fit is a good approximation for H ionizing photons. While we find that a blackbody fitting overestimates He~II photons at initial masses $\leq 40\,\msun$, this is unexpected to impact our ionizing photon production results for the population since the He~II photon production for all IMF slopes $\leq 2.35$ is dominated by stars with $\mini > 40\,\msun$.
    \item Through comparing our results to \cite{Yoon2012EvolutionFields}, we have found that models that achieve CHE produce significantly more ionizing photons with increases of up to 1.4\,dex. This shows that massive rotators should produce more ionizing photons if magnetic fields are included, which will impact top-heavy IMF populations significantly.
\end{itemize}


\section*{Acknowledgements} We wish to acknowledge the Irish Research Council (IRC) for funding this research, as well as "ChETEC" COST Action (CA16117), supported by COST (European Cooperation in Science and Technology) which aided our collaboration with co-authors. Georges Meynet and Sylvia Ekstr\"om have received funding from the European Research Council (ERC) under the European Union's Horizon 2020 research and innovation programme (grant agreement no. 833925, project STAREX).

\section*{Data availability} The derived data generated in this research will be shared on reasonable request to the corresponding author.

\bibliography{refpapers}
\bibliographystyle{mnras}

\label{lastpage}
\end{document}